\documentclass[12pt]{article}
\usepackage{amsmath}

\newcommand{\eq}{\begin{equation}}
\newcommand{\qe}{\end{equation}}
\newcommand{\eqa}{\begin{eqnarray*}}
\newcommand{\qa}{\end{eqnarray*}}

\begin{document}

\title{Scattering in $\mathcal{P}\mathcal{T}$-Symmetric Quantum Mechanics }
\author{F. Cannata \\
Istituto Nazionale di Fisica Nucleare, Sezione di Bologna\\
and Dipartimento di Fisica dell' Universit\`a,\\
Via Irnerio 46, I 40126 Bologna, Italy \\
\\
J.-P. Dedonder\\
GMPIB Universit\'e Paris 7 - Denis-Diderot \\
2 Place Jussieu,\\
F-75251, Paris Cedex 05, France \\
\\
A. Ventura\\
Ente Nuove Tecnologie, Energia e Ambiente, Bologna\\
and Istituto Nazionale di Fisica Nucleare, Sezione di Bologna, Italy}
\maketitle

\begin{abstract}
A general formalism is worked out for the description of one-dimensional
scattering in non-hermitian quantum mechanics and constraints on
transmission and reflection coefficients are derived in the cases of $%
\mathcal{P}$, $\mathcal{T}$ or $\mathcal{PT}$ invariance of the Hamiltonian.
Applications to some solvable $\mathcal{PT}$-symmetric potentials are shown
in detail.

Our main original results concern the association of reflectionless
potentials with asymptotic exact $\mathcal{PT}$ symmetry and the
peculiarities of separable kernels of non-local potentials in connection
with Hermiticity, $\mathcal{T}$ invariance and $\mathcal{PT}$ invariance. 
\end{abstract}

\vspace{.5cm}

\newpage

\section{Introduction}

Since the seminal paper by Bender and Boettcher~\cite{BB98}, research on $%
\mathcal{PT}$-symmetric quantum mechanics has been mainly focused on bound
states, either with real energies, or in complex conjugate energy pairs,
while relatively few authors have studied scattering states of Hamiltonians
with both discrete and continuous spectra\cite{An99, Le01, Le02, De03}.

This gap has been recently bridged, at least in part, by a review paper on
complex absorbing potentials~\cite{Mu04}, covering the more extended topics
of one-dimensional scattering in non-hermitian quantum mechanics, with some
general formulae valid for complex $\mathcal{PT}$-symmetric potentials.
Nonetheless, we think it worthwhile to go into further details in the latter
case, while paying the due credit to the authors of Ref.~\cite{Mu04}, and
apply our formalism to some examples of solvable potentials for the sake of
clarity.

In any case, we share the philosophy of ref.~\cite{Mu04}, i. e. to consider $%
\mathcal{PT}$-symmetric potentials as complex potentials defined on the real
axis. Therefore, when we define complex coordinate shifts, we mean this as a
method of generating complex potentials depending on a real coordinate.

The main purpose of the present paper is thus to give a general description
of one-dimensional scattering in non-hermitian quantum mechanics with the
Hamiltonian, $H$, invariant either under parity, $\mathcal{P}$, or time
reversal, $\mathcal{T}$, or their product, $\mathcal{PT}$, in the case $H$
is not separately invariant under $\mathcal{P}$ and $\mathcal{T}$. We do not
intend, however, to provide here physical motivations for $\mathcal{PT}$
symmetry, nor to elaborate on the significance of violation of hermiticity
and unitarity in quantum mechanics. The reader interested in these topics
can consult a wide list of references, in particular the recent works~\cite
{Mo04, We04, Ru05} and references therein.

In the present work, we limit ourselves to a framework in which the
scattering potentials vanish asymptotically: in particular, we exclude
potentials diverging at infinity, like those considered in Ref.\cite{ABB05}.
Therefore, our asymptotic wave functions are linear superpositions of plane
waves and the present approach closely resembles standard scattering theory
described in text-books on quantum mechanics, such as Ref.\cite{Me98}, in
the limit in which Hermiticity and $\mathcal{P-}$ and $\mathcal{T-}$%
symmetries hold.

What we try to achieve is to assemble a comprehensive and self-contained
formalism for discussing scattering problems in one-dimensional quantum
mechanics. With this formalism available, we provide suitable examples,
which should make the reader capable of elaborating his/her judgement on the
relevance of the subject of $\mathcal{PT}$ symmetry, including some
contributions to a better understanding of ``exact'' $\mathcal{PT}$
symmetry. Due to the relation of the Schr\"{o}dinger equation to the
classical Helmholtz equation, our formalism may be accommodated to deal with
optics~\cite{Mu04, CW94} with refraction index characterized by
``handedness''~\cite{Ah04}.

In addition, we try to clarify the interplay between $\mathcal{T}$
invariance and hermiticity, displaying a complex non-local solvable
potential in which hermiticity does not force $\mathcal{T}$ invariance, but
can be compatible with $\mathcal{PT}$ invariance. To our knowledge, a $%
\mathcal{PT-}$symmetric non-local potential is introduced and worked out
here for the first time.

We hope that our work is sufficiently self-contained, such as not to require
any particular specific background of the reader. The paper definitely has
topical review aspects, though it does not pretend to give a complete list
of references. There are, however, relevant original results, in particular
for non-local separable potentials.

The paper is organized as follows: section~\ref{L-R} describes the basic
formalism of one-dimensional scattering in non-hermitian quantum mechanics,
section~\ref{symm} introduces symmetries under which $H$ may be invariant,
section~\ref{curr} defines density currents and continuity equations,
section~\ref{examples} applies the formalism worked out in the previous
three sections to some solvable potentials. Finally, section~\ref{concl} is
devoted to conclusions.

\section{\textit{L}-\textit{R} Representation}

\label{L-R}

We start from the general time-dependent Schr\"{o}dinger equation 
\begin{equation}
-\frac{\partial ^2}{\partial x^2}\psi \left( x,t\right) +\int K\left(
x,y\right) \psi \left( y,t\right) dy=i\frac \partial {\partial t}\psi \left(
x,t\right) ,  \label{Schreqt}
\end{equation}
written in units $\hbar =2m=1$. For a monochromatic wave, of energy $\omega $%
, the time dependence of the wave function is 
\begin{equation}
\psi \left( x,t\right) =\Psi \left( x\right) e^{-i\omega t}\,.  \label{psi_t}
\end{equation}
In the present work, unless otherwise stated, we consider local potentials,
for which the kernel $K$ reduces to 
\begin{equation}
K\left( x,y\right) =\delta \left( x-y\right) V\left( y\right)   \label{local}
\end{equation}
If eqs.(\ref{psi_t}-\ref{local}) hold, eq. (\ref{Schreqt}) reduces to the
time independent Schr\"{o}dinger equation satisfied by $\Psi (x)$ 
\begin{equation}
H\Psi \equiv \left( -{\frac{d^2}{dx^2}}+V(x)\right) \Psi =k^2\Psi ,
\label{Schreq}
\end{equation}
with $k=\sqrt{\omega }$ ($>0$) the wave number. In order to solve eq.(\ref
{Schreq}), it is convenient to work in a two dimensional Hilbert space where
the basis vectors are the kets $|R>$ and $|L>$ (and the corresponding bras $%
<R|$ and $<L|$). In configuration space,with the choice of the time
dependent phase given by eq. (\ref{psi_t}), 
\begin{equation}
<x|R,k>\,\sim \,e^{ikx}
\end{equation}
represents a plane wave travelling from left to right and 
\begin{equation}
<x|L,k>\,\sim \,e^{-ikx}
\end{equation}
a wave travelling from right to left. In the following, the explicit $k$
dependence of the basis vectors will be omitted, whenever not strictly
necessary, for simplicity of notation.

In coordinate space, the asymptotic states , i.e., outside the (finite)
range of the potential, can be expressed at $x\rightarrow -\infty $ as 
\begin{equation}
\left| \Psi _{x\rightarrow -\infty }\right\rangle =A_{-}|R>+B_{-}|L>
\end{equation}
and, at $x\rightarrow +\infty $, as 
\begin{equation}
\left| \Psi _{x\rightarrow +\infty }\right\rangle =A_{+}|R>+B_{+}|L>,
\end{equation}
or, in terms of wave functions 
\begin{equation}
\Psi (x)=\left\{ 
\begin{array}{lcr}
A_{-}e^{ikx}+B_{-}e^{-ikx} &  & \mbox{$x \rightarrow -\infty$} \\ 
A_{+}e^{ikx}+B_{+}e^{-ikx} &  & \mbox{$x \rightarrow +\infty$}
\end{array}
\right.  \label{Progwave}
\end{equation}

In the case of a finite-range local potential, Eq. (\ref{Schreq}) admits a
general solution written as a linear combination of two independent
solutions, $F_1\left( x\right) $ and $F_2\left( x\right) $, with non-zero
Wronskian, whose asymptotic expressions are both of the form: 
\begin{equation}
\lim_{x\rightarrow \pm \infty }F_m\left( x\right) =a_{m\pm }e^{ikx}+b_{m\pm
}e^{-ikx},\qquad (m=1,2)  \label{F_i}
\end{equation}
The $a_{m\pm }$ and $b_{m\pm }$ are simply related to the asymptotic
amplitudes $A_{\pm }$ and $B_{\pm }$ 
\begin{eqnarray}
A_{\pm } &=&\alpha a_{1\pm }+\beta a_{2\pm }  \label{XY} \\
B_{\pm } &=&\alpha b_{1\pm }+\beta b_{2\pm }\,.  \nonumber
\end{eqnarray}
If $\Psi _1(x)=\alpha F_1\left( x\right) +\beta F_2\left( x\right) $ ,
inserted into Eq. (\ref{psi_t}), gives rise to a wave moving from $x=-\infty $
to $x=+\infty $, the amplitude of the regressive wave vanishes at $+\infty $%
: 
\[
B_{+}=0\Longrightarrow \frac \beta \alpha =-\frac{b_{1+}}{b_{2+}}, 
\]
and the transmission and reflection coefficients of the wave moving from
left to right are immediately written as 
\begin{eqnarray*}
T_{L\rightarrow R}\equiv \frac{A_{+}}{A_{-}} &=&\frac{\alpha a_{1+}+\beta
a_{2+}}{\alpha a_{1-}+\beta a_{2-}} \\
&=&\frac{a_{2+}b_{1+}-a_{1+}b_{2+}}{a_{2-}b_{1+}-a_{1-}b_{2+}}.
\end{eqnarray*}
\begin{eqnarray*}
R_{L\rightarrow R}\equiv \frac{B_{-}}{A_{-}} &=&\frac{\alpha b_{1-}+\beta
b_{2-}}{\alpha a_{1-}+\beta a_{2-}} \\
&=&\frac{b_{1+}b_{2-}-b_{1-}b_{2+}}{a_{2-}b_{1+}-a_{1-}b_{2+}}.
\end{eqnarray*}
The asymptotic form of $\Psi _1(x)$ is thus, neglecting a global
normalization factor 
\begin{eqnarray}
\Psi _1(x) &\sim &e^{ikx}+R_{L\rightarrow R}e^{-ikx},\quad x\rightarrow
-\infty  \label{Psi_1} \\
&\sim &T_{L\rightarrow R}e^{ikx},\quad x\rightarrow +\infty \,.  \nonumber
\end{eqnarray}

Similarly, if $\Psi _2\left( x\right) $ gives rise to a wave moving from $%
x=+\infty $ to $x=-\infty $, the amplitude of the progressive wave vanishes
at $-\infty $: 
\[
\widetilde{A}_{-}=0\Longrightarrow \frac \beta \alpha =-\frac{a_{1-}}{a_{2-}}%
, 
\]
and the transmission and reflection coefficients of the wave moving from
right to left are: 
\begin{eqnarray*}
T_{R\rightarrow L}\equiv \frac{\widetilde{B}_{-}}{\widetilde{B}_{+}} &=&%
\frac{\alpha b_{1-}+\beta b_{2-}}{\alpha b_{1+}+\beta b_{2+}} \\
&=&\frac{a_{2-}b_{1-}-a_{1-}b_{2-}}{a_{2-}b_{1+}-a_{1-}b_{2+}}.
\end{eqnarray*}
\begin{eqnarray*}
R_{R\rightarrow L}\equiv \frac{\widetilde{A}_{+}}{\widetilde{B}_{+}} &=&%
\frac{\alpha a_{1+}+\beta a_{2+}}{\alpha b_{1+}+\beta b_{2+}} \\
&=&\frac{a_{1+}a_{2-}-a_{1-}a_{2+}}{a_{2-}b_{1+}-a_{1-}b_{2+}}.
\end{eqnarray*}
As a consequence, the asymptotic form of $\Psi _2(x)$ is 
\begin{eqnarray}
\Psi _2(x) &\sim &T_{R\rightarrow L}e^{-ikx},\quad x\rightarrow -\infty
\label{Psi_2} \\
&\sim &e^{-ikx}+R_{R\rightarrow L}e^{ikx},\quad x\rightarrow +\infty 
\nonumber
\end{eqnarray}
When we compute the Wronskian of $\Psi _1$ and $\Psi _2$, defined as 
\[
W(x)=\Psi _1(x)\frac d{dx}\Psi _2(x)-\Psi _2(x)\frac d{dx}\Psi _1(x), 
\]
we readily obtain $W(-\infty )=-2ikT_{R\rightarrow L}$ and $W(+\infty
)=-2ikT_{L\rightarrow R}$. Therefore, a necessary condition for the
Wronskian to be constant on the $x$ axis is $T_{L\rightarrow
R}=T_{R\rightarrow L}$. It is easy to check that ${dW}/{dx}=0$ for any
well-behaved local potential. Therefore, the equality of the two
transmission coefficients is satisfied for any such potential.

The scattering matrix, $S$, connects the outgoing states at $t\rightarrow
+\infty $ to the ingoing ones at $t\rightarrow -\infty $ 
\begin{equation}
\left| \Psi _{out}\right\rangle =S\left| \Psi _{in}\right\rangle \,.
\end{equation}

The $S$ matrix elements are directly linked to the transmission and
reflection coefficients. In this basis, $|R>$ and $|L>$ can be rewritten as 
\[
\left| R\right\rangle \equiv \left( 
\begin{array}{c}
1 \\ 
0
\end{array}
\right) 
\]
and 
\[
\left| L\right\rangle \equiv \left( 
\begin{array}{c}
0 \\ 
1
\end{array}
\right) \,. 
\]
If we have an ingoing wave of $\left| R\right\rangle $ type, then the
outgoing wave is \newline
\[
\left( 
\begin{array}{l}
A_{+} \\ 
B_{-}
\end{array}
\right) =S\left( 
\begin{array}{l}
A_{-} \\ 
0
\end{array}
\right) \Longrightarrow \left( 
\begin{array}{c}
T_{L\rightarrow R} \\ 
R_{L\rightarrow R}
\end{array}
\right) =S\left( 
\begin{array}{c}
1 \\ 
0
\end{array}
\right) . 
\]
\newline
If the ingoing wave is of $\left| L\right\rangle $ type, we obtain, with the
same procedure 
\begin{equation}
\left( 
\begin{array}{c}
R_{R\rightarrow L} \\ 
T_{R\rightarrow L}
\end{array}
\right) =S\left( 
\begin{array}{c}
0 \\ 
1
\end{array}
\right) .
\end{equation}
As a consequence of the above equations, the $S$ matrix elements are
explicitly given by 
\begin{equation}
S=\left( 
\begin{array}{lr}
S_{RR} & S_{RL} \\ 
S_{LR} & S_{LL}
\end{array}
\right) =\left( 
\begin{array}{lr}
T_{L\rightarrow R} & R_{R\rightarrow L} \\ 
R_{L\rightarrow R} & T_{R\rightarrow L}
\end{array}
\right) .  \label{Smat}
\end{equation}

Note that our $T_{i\rightarrow j}(R_{i\rightarrow j})$ corresponds to $%
T^i(R^i)$ of Ref.~\cite{Mu04} and to $t_j(r_j)$ of Ref.~\cite{De03}. Our
definition of $S$ matrix is the same as that of Refs.~\cite{Mu04, De03} ,
while in the $S$ matrix defined in the book by Merzbacher~\cite{Me98} the
rows are exchanged with respect to ours. The latter author introduces also
the transfer matrix, $M$, yielding the asymptotic states at $x\rightarrow
-\infty $ when applied to those at $x\rightarrow +\infty $ (see Ref.\cite
{Me98}, formula (6.24)). In our basis, $M$, like $S$, is a $2\times 2$
matrix, whose elements are easily expressed in terms of right and left
transmission and reflection coefficients. 
\begin{equation}
\left( 
\begin{array}{c}
A_{-} \\ 
B_{-}
\end{array}
\right) =M\left( 
\begin{array}{c}
A_{+} \\ 
B_{+}
\end{array}
\right) .
\end{equation}

If the incident wave is of $\left|R\right\rangle$ type, the effect of $M$ is
expressed by the equation \newline
\[
\left( 
\begin{array}{c}
1 \\ 
R_{L\rightarrow R}
\end{array}
\right) = M \left( 
\begin{array}{c}
T_{L\rightarrow R} \\ 
0
\end{array}
\right), 
\]
while, if the incident wave is of $\left|L\right\rangle$ type, the following
equation holds 
\[
\left( 
\begin{array}{c}
0 \\ 
T_{R\rightarrow L}
\end{array}
\right)= M\left( 
\begin{array}{c}
R_{R\rightarrow L} \\ 
1
\end{array}
\right). 
\]

From the equations given above, we obtain 
\begin{equation}
\left( 
\begin{array}{ll}
M_{RR} & M_{RL} \\ 
M_{LR} & M_{LL}
\end{array}
\right) =\left( 
\begin{array}{cc}
1/T_{L\rightarrow R} & \qquad -R_{R\rightarrow L}/T_{L\rightarrow R} \\ 
R_{L\rightarrow R}/T_{L\rightarrow R} & \qquad T_{R\rightarrow
L}-R_{R\rightarrow L}R_{L\rightarrow R}/T_{L\rightarrow R}
\end{array}
\right) ,  \label{M_mat}
\end{equation}
with $\det M=T_{R\rightarrow L}/T_{L\rightarrow R}$.

In general, any matrix, $O$, in the $R-L$ basis 
\begin{equation}
O\equiv \left( 
\begin{array}{ll}
O_{RR} & O_{RL} \\ 
O_{LR} & O_{LL}
\end{array}
\right)
\end{equation}
\newline
can be written also as a linear combination of basic dyadic operators 
\begin{equation}
\Omega _{ij}=|i><j|\;,\qquad j=R,L
\end{equation}
in the form 
\begin{equation}
O=O_{RR}\Omega _{RR}+O_{RL}\Omega _{RL}+O_{LR}\Omega _{LR}+O_{LL}\Omega _{LL}
\end{equation}

This notation will be used in the following sections in the discussion of
the invariance of the Hamiltonian with respect to various transformations.

\section{$\mathcal{P}$, $\mathcal{T}$ and $\mathcal{PT}$ Symmetries}

\label{symm}

In the present section, we study the transformation properties of the
Hamiltonian with respect to $\mathcal{P}$, $\mathcal{T}$ and $\mathcal{PT}$
reflections. With particular reference to $\mathcal{PT}$ transformations, it
is useful to introduce a discussion of the behaviour of vectors and matrices
in the $R-L$ basis under coordinate shifts.

\subsection{ Coordinate Shifts}

\label{shifts}

In connection with the coordinate shift $x\rightarrow x+X_0$, with $X_0$ a
real number,

let us define a displacement operator, $\mathcal{D}(X_0)$, through its
action on the basis vectors $\left| R\right\rangle $ and $\left|
L\right\rangle $%
\begin{eqnarray*}
\mathcal{D}(X_0)\left| R\right\rangle &=&e^{ikX_0}\left| R\right\rangle \,;
\\
\mathcal{D}(X_0)\left| L\right\rangle &=&e^{-ikX_0}\left| L\right\rangle \,;
\end{eqnarray*}
and on their dual vectors 
\begin{eqnarray*}
\left\langle R\right| \mathcal{D}^{-1}(X_0) &=&e^{-ikX_0}\left\langle
R\right| ; \\
\left\langle L\right| \mathcal{D}^{-1}(X_0) &=&e^{ikX_0}\left\langle
L\right| .
\end{eqnarray*}
In matrix form 
\begin{equation}
\mathcal{D}(X_0)=\left( 
\begin{array}{cc}
e^{ikX_0} & 0 \\ 
0 & e^{-ikX_0}
\end{array}
\right) .  \label{Shift}
\end{equation}

Since $X_0$ is real, ${\mathcal{D}}^{-1}(X_0)=\mathcal{D}^{*}(X_0)$.

The basic dyadic operators are thus transformed under $D$ according to the
obvious relations 
\begin{eqnarray*}
\mathcal{D}(X_0)\left| R\right\rangle \left\langle R\right| \mathcal{D}%
^{-1}(X_0) &=&\left| R\right\rangle \left\langle R\right| ; \\
\mathcal{D}(X_0)\left| L\right\rangle \left\langle L\right| \mathcal{D}%
^{-1}(X_0) &=&\left| L\right\rangle \left\langle L\right| ; \\
\mathcal{D}(X_0)\left| R\right\rangle \left\langle L\right| \mathcal{D}%
^{-1}(X_0) &=&e^{2ikX_0}\left| R\right\rangle \left\langle L\right| ; \\
\mathcal{D}(X_0)\left| L\right\rangle \left\langle R\right| \mathcal{D}%
^{-1}(X_0) &=&e^{-2ikX_0}\left| L\right\rangle \left\langle R\right| .
\end{eqnarray*}
and a generic one-body operator, conveniently written in the form of a $%
2\times 2$ matrix, is consequently transformed as follows 
\begin{equation}
\mathcal{D}(X_0)O\mathcal{D}^{-1}(X_0)=\mathcal{D}(X_0)\left( 
\begin{array}{ll}
O_{RR} & O_{RL} \\ 
O_{LR} & O_{LL}
\end{array}
\right) \mathcal{D}^{-1}(X_0)=\left( 
\begin{array}{ll}
O_{RR} & e^{2ikX_0}O_{RL} \\ 
e^{-2ikX_0}O_{LR} & O_{LL}
\end{array}
\right) .  \label{C_shift}
\end{equation}

\subsection{Parity}

\label{par} We shall consider now the parity transformation 
\[
x \rightarrow - x, \qquad p_x\rightarrow -p_x\,. 
\]
Under parity, the kets $|R> $ and $|L>$ are transformed according to 
\[
\begin{array}{l}
\mathcal{P} |R> = |L> ; \\ 
\mathcal{P} |L> = |R>,
\end{array}
\]
and the bras $< R |$ and $< L |$ are changed as follows 
\[
\begin{array}{l}
< R | \mathcal{P}^{-1} = < L |; \\ 
< L | \mathcal{P}^{-1} = < R |.
\end{array}
. 
\]

Hence, the four basic operators $\Omega_{ij} $ transformed according to $%
\mathcal{P} \Omega_{ij} \mathcal{P}^{-1}$ yield 
\begin{equation}
\begin{array}{lr}
\Omega_{RR} & \Omega_{RL} \\ 
\Omega_{LR} & \Omega_{LL}
\end{array}
\rightarrow 
\begin{array}{lr}
\Omega_{LL} & \Omega_{LR} \\ 
\Omega_{RL} & \Omega_{RR}
\end{array}
.  \label{Qtransf}
\end{equation}

In this basis, the parity operator can be represented by the matrix $%
\mathcal{P}$ ( with $\mathcal{P}^2 = 1$)~\cite{De03}, which, acting on the
left, exchanges lines, while acting on the right exchanges columns 
\begin{equation}
\mathcal{P} = \left( 
\begin{array}{lr}
0 & 1 \\ 
1 & 0
\end{array}
\right) = \mathcal{P}^{-1}  \label{P0}
\end{equation}
\newline
The Hamiltonian can be expressed as 
\begin{equation}
H \equiv H_{RR} \Omega_{RR} + H_{RL} \Omega_{RL} + H_{LR} \Omega_{LR} +
H_{LL} \Omega_{LL} ,  \label{hamilton}
\end{equation}
or, in matrix form 
\[
H = \left( 
\begin{array}{lr}
H_{RR} & H_{RL} \\ 
H_{LR} & H_{LL}
\end{array}
\right). 
\]
The transformed Hamiltonian is then 
\begin{eqnarray*}
H_{P} & = & \mathcal{P} H \mathcal{P}^{-1} \\
& = & H_{RR} \Omega_{LL} + H_{RL} \Omega_{LR} + H_{LR} \Omega_{RL} + H_{LL}
\Omega_{RR}  \nonumber \\
& = & \left( 
\begin{array}{lc}
H_{LL} & H_{LR} \\ 
H_{RL} & H_{RR}
\end{array}
\right) .  \nonumber
\end{eqnarray*}

Parity invariance for the Hamiltonian, $H_{\mathcal{P}}=H$, therefore
requires 
\begin{equation}
H_{RR}=H_{LL}  \label{PHinva}
\end{equation}
\begin{equation}
H_{RL}=H_{LR}.  \label{PHinvb}
\end{equation}

In the interaction picture, the $S$ matrix is known to be the following
limit of the transition operator, $\widetilde{T}(t-t_0)$ (see Ref.\cite{Me98}%
, formulae (14.49) and (20.7) ) 
\begin{equation}
S=\lim_{t\rightarrow \infty ,t_0\rightarrow -\infty }\widetilde{T}%
(t-t_0)=\lim_{t\rightarrow \infty ,t_0\rightarrow -\infty }U_0\left(
t\right) T(t-t_0)U_0^{-1}\left( t_0\right)  \label{TrOp}
\end{equation}
with 
\begin{equation}
U_0\left( t\right) =e^{iH_0t}\;,\;H_0=p^2=-\frac{d^2}{dx^2}
\end{equation}
and 
\begin{equation}
T(t-t_0)=e^{-iH(t-t_0)}=\sum_{n=0}^\infty (-i)^nH^n(t-t_0)^n/n!
\label{opevol}
\end{equation}

It is now possible to investigate the transformation of $S$ under parity $%
\mathcal{P}$. In order to compute 
\begin{equation}
S_{\mathcal{P}}=\mathcal{P}S\mathcal{P}^{-1}
\end{equation}
we need 
\begin{eqnarray*}
\mathcal{P}T(t-t_0)\mathcal{P}^{-1} &=&\sum_{n=0}^\infty (-i)^n\mathcal{P}H^n%
\mathcal{P}^{-1}(t-t_0)^n/n! \\
&=&\sum_{n=0}^\infty (-i)^nH_{\mathcal{P}}^n(t-t_0)^n/n! \\
&=&\sum_{n=0}^\infty (-i)^nH^n(t-t_0)^n/n!=T(t-t_0)\,,
\end{eqnarray*}
and $\mathcal{P}U_0\left( t\right) \mathcal{P}^{-1}=U_0\left( t\right) $, so
that 
\begin{equation}
S_{\mathcal{P}}=\mathcal{S}.
\end{equation}
Hence, the $S$ matrix commutes with $\mathcal{P}$.

Therefore, if the Hamiltonian is invariant under parity, so is the $S$
matrix, which yields, upon using eq. (\ref{Smat}) 
\begin{equation}
S_{RR} = S_{LL} \rightarrow T_{L\rightarrow R}=T_{R\rightarrow L}
\end{equation}
and 
\begin{equation}
S_{RL} = S_{LR} \rightarrow R_{R\rightarrow L}=R_{L\rightarrow R}
\end{equation}

The last two equations are strictly analogous to those, eqs.(\ref{PHinva})
and (\ref{PHinvb}), expressing the parity invariance of the Hamiltonian, and
are already given in Ref.~\cite{Mu04}.

\subsection{Generalized Parity}

\label{gpar}

We shall consider the following generalized parity transformation 
\[
x\rightarrow X_0-x\,\quad X_0\in \mathbf{R} 
\]
corresponding to the reflection around the point $X_0/2$ on the real x axis.
This kind of transformation will be useful in the study of properties of
potentials not centred on the origin, and can be readily expressed as the
(non commutative) product of the parity operator $\mathcal{P}$ and the
coordinate shift operator $\mathcal{D}(X_0)$ 
\begin{equation}
\mathcal{P}_G(X_0)\equiv \mathcal{P}\mathcal{D}(X_0).  \label{P_G}
\end{equation}
Under generalized parity, the kets $|R>$ and $|L>$ are transformed as
follows 
\[
\begin{array}{l}
\mathcal{P}_G(X_0)|R>=e^{ikX_0}|L> \\ 
\mathcal{P}_G(X_0)|L>=e^{-ikX_0}|R>
\end{array}
\]
and, correspondingly, the bras $<R|$ and $<L|$ give 
\[
\begin{array}{l}
<R|\mathcal{P}_G^{-1}(X_0)=<L|e^{-ikX_0} \\ 
<L|\mathcal{P}_G^{-1}(X_0)=<R|e^{ikX_0}
\end{array}
\]
Since the matrix $\mathcal{P}_G(X_0)$ associated to this transformation
reads in the $|R>,|L>$ basis 
\[
\mathcal{P}_G(X_0)=\left( 
\begin{array}{lr}
0 & e^{-ikX_0} \\ 
e^{ikX_0} & 0
\end{array}
\right) \label{PX0} 
\]
and, consequently, 
\[
\mathcal{P}_G(X_0)=\mathcal{P}_G^{-1}(X_0), 
\]
\newline
the four basic operators $\Omega _{ij}$ transformed under $\mathcal{P}%
_G(X_0)\Omega _{ij}\mathcal{P}_G^{-1}(X_0)$ yield

\begin{equation}
\begin{array}{lr}
\Omega _{RR} & \Omega _{RL} \\ 
\Omega _{LR} & \Omega _{LL}
\end{array}
\rightarrow 
\begin{array}{lr}
\Omega _{LL} & \Omega _{LR}e^{2ikX_0} \\ 
\Omega _{RL}e^{-2ikX_0} & \Omega _{RR}
\end{array}
\end{equation}

Thus, the transformed Hamiltonian reads 
\[
\begin{array}{ll}
H_{\mathcal{P}_G}=\mathcal{P}_G(X_0)H\mathcal{P}_G^{-1}(X_0) &  \\ 
=H_{RR}\Omega _{LL}+H_{RL}\Omega _{LR}e^{2ikX_0}+H_{LR}\Omega
_{RL}e^{-2ikX_0}+H_{LL}\Omega _{RR} & 
\end{array}
\]
Invariance of the Hamiltonian under generalized parity, $H_{\mathcal{P}_G}=H$%
, is equivalent to 
\begin{equation}
H_{RR}=H_{LL}  \label{PGHinva}
\end{equation}
\begin{equation}
H_{RL}=H_{LR}e^{-2ikX_0}.  \label{PGHinvb}
\end{equation}
The behaviour of the $S$ matrix under $\mathcal{P}_G$ immediately follows
from the $\mathcal{P}_G$ invariance of $H$ 
\[
S_{\mathcal{P}}=\mathcal{P}_G(X_0)S\mathcal{P}_G^{-1}(X_0) 
\]
We have 
\[
S_{\mathcal{P}_G}=S\,, 
\]
or 
\[
\mathcal{P}_G(X_0)S=S\mathcal{P}_G(X_0)\,. 
\]
$\mathcal{P}_G$ invariance of the $S$ matrix corresponds to 
\[
S_{RR}=S_{LL}\,\rightarrow \,T_{R\rightarrow L}=T_{L\rightarrow L} 
\]
and 
\[
S_{RL}e^{ikX_0}=S_{LR}e^{-ikX_0}\,\rightarrow \,R_{R\rightarrow
L}e^{ikX_0}=R_{L\rightarrow R}e^{-ikX_0}. 
\]

\subsection{Time Reversal}

\label{tr}

For this case we follow the same steps as in subsection~\ref{par}. First, we
consider the consequence of the invariance property for the Hamiltonian (\ref
{hamilton}), having in mind that the time reversal operator $\mathcal{T}$ is
an antiunitary operator ($\mathcal{T} = \mathcal{T}^{-1} = \mathcal{T}%
^{\dag} $). Under time reversal 
\[
x\rightarrow x,\quad p_x\rightarrow -p_x\quad i\rightarrow -i \nonumber
\]
any ket in the $|R>$, $|L>$ basis is transformed according to 
\[
\mathcal{T} (\alpha |R> + \beta |L>) = \alpha^{*} |L> +\beta^{*} |R>, 
\]
while a bra fulfills the relation 
\[
(\gamma < R | + \delta < L |) \mathcal{T}^{-1} = \gamma^{*} < L | +
\delta^{*} < R | 
\]

It is worth stressing that $\mathcal{T}$-invariance as such does not force
the potential to be real, and hence, the Hamiltonian to be hermitian.

The $\Omega _{ij}$ operators are now transformed into $\mathcal{T}\Omega
_{ij}\mathcal{T}^{-1}$ 
\begin{equation}
\begin{array}{lr}
\Omega _{RR} & \Omega _{RL} \\ 
\Omega _{LR} & \Omega _{LL}
\end{array}
\rightarrow 
\begin{array}{lr}
\Omega _{LL} & \Omega _{LR} \\ 
\Omega _{RL} & \Omega _{RR}
\end{array}
\label{Ttransf}
\end{equation}
which may be compared to eqs.(\ref{Qtransf}) and (\ref{P_G}). The $\mathcal{T%
}-$transformed Hamiltonian, $H_{\mathcal{T}}=\mathcal{THT}^{-1}$, reads 
\begin{equation}
H_{\mathcal{T}}=H_{RR}^{*}\Omega _{LL}+H_{RL}^{*}\Omega
_{LR}+H_{LR}^{*}\Omega _{RL}+H_{LL}^{*}\Omega _{RR}=\left( 
\begin{array}{lr}
H_{LL}^{*} & H_{LR}^{*} \\ 
H_{RL}^{*} & H_{RR}^{*}
\end{array}
\right) \,.  \label{H_T}
\end{equation}

From the properties of the parity operator (\ref{P0}), it is easy to realize
that $H_{\mathcal{T}}$ defined in the previous equation is such that 
\begin{equation}
\mathcal{P}H_{\mathcal{T}}=H^{*}\mathcal{P}\,,  \label{HTtransf}
\end{equation}
or 
\begin{equation}
H_{\mathcal{T}}=\mathcal{P}H^{*}\mathcal{P}\,,  \label{HT1}
\end{equation}
where $\mathcal{P}$ is the parity operator (\ref{P0}).

We now proceed to examine how the $S$ matrix transforms under $\mathcal{T}$
without imposing any symmetry to the Hamiltonian. Starting from the
transition operator, $\widetilde{T}(t-t_0)$, of formula (\ref{TrOp}), and
using eq. (\ref{HT1}), we have 
\begin{eqnarray*}
\widetilde{T}_{\mathcal{T}}(t-t_0) &\equiv &\mathcal{T}\widetilde{T}(t-t_0)%
\mathcal{T}^{-1} \\
&=&\mathcal{T}e^{iH_0t}\mathcal{T}^{-1}\sum_{n=0}^\infty \mathcal{T}(-i)^nH^n%
\mathcal{T}^{-1}(t-t_0)^n/n!\mathcal{T}e^{-iH_0t_0}\mathcal{T}^{-1} \\
&=&e^{-iH_0t}\sum_{n=0}^\infty i^nH_{\mathcal{T}}^n(t-t_0)^n/n!e^{iH_0t_0} \\
&=&e^{-iH_0t}\sum_{n=0}^\infty i^n(\mathcal{P}H^{*}\mathcal{P})^n(t-t_0)^n/n!e^{iH_0t_0}
={\mathcal{P}}\widetilde{T}^{*}(t-t_0)\mathcal{P},
\end{eqnarray*}
which, when $t\rightarrow +\infty $ and $t_0\rightarrow -\infty $, yields 
\begin{equation}
S_{\mathcal{T}}\equiv {\mathcal{T}}S{\mathcal{T}}^{-1}={\mathcal{P}}S^{*}\mathcal{P%
}=\left( 
\begin{array}{cc}
S_{LL}^{*} & S_{LR}^{*} \\ 
S_{RL}^{*} & S_{RR}^{*}
\end{array}
\right) .  \label{S_T}
\end{equation}

\subsubsection{$H_{\mathcal{T}}=H$}

\label{T_inv} For a $\mathcal{T}$-invariant Hamiltonian ($H_{\mathcal{T}} =H$%
), the diagonal matrix elements are complex conjugate of each other as are
the two non-diagonal ones, so that a time reversal invariant Hamiltonian is,
in that basis, of the form 
\begin{equation}
H = H_{\mathcal{T}} = \left( 
\begin{array}{lr}
H_{RR} & H_{RL} \\ 
H_{RL}^{*} & H_{RR}^{*}
\end{array}
\right)\,.  \label{HTtransf1}
\end{equation}
Hence, as is well-known, this is not equivalent to hermiticity of the
Hamiltonian.

Coming now to the transition operator, $T$, we easily obtain, under time
reversal invariance of $H$, with the same procedure as in the previous
subsection, 
\[
T_{\mathcal{T}}(t-t_0)=T^{-1}(t-t_0)\,. 
\]
Hence, $H=H_{\mathcal{T}}$ leads to 
\begin{equation}
S_{\mathcal{T}}=S^{-1}\equiv {\frac 1{\det S}}\left( 
\begin{array}{cc}
S_{LL} & -S_{RL} \\ 
-S_{LR} & S_{RR}
\end{array}
\right) ,  \label{S_T_inv}
\end{equation}

Simultaneous validity of Eqs. (\ref{S_T}, \ref{S_T_inv}) thus leads to the
following explicit relations for the S-matrix elements : 
\begin{eqnarray*}
S_{RL}+S_{LR}^{*}\det S &=&0 \\
S_{LR}+S_{RL}^{*}\det S &=&0 \\
S_{LL} &=&S_{LL}^{*}\det S \\
S_{RR} &=&S_{RR}^{*}\det S\,.
\end{eqnarray*}
Hence, on the assumption that all the $S$-matrix elements are different from
zero, we obtain 
\[
\det S=\frac{S_{LL}}{S_{LL}^{*}}=\frac{S_{RR}}{S_{RR}^{*}}=-\frac{S_{RL}}{%
S_{LR}^{*}}=-\frac{S_{LR}}{S_{RL}^{*}}\,, 
\]
with 
\[
\mid \det S\mid =1 
\]
and 
\[
\mid S_{LR}\mid =\mid S_{RL}\mid 
\]
or, equivalently 
\[
\mid R_{L\rightarrow R}\mid =\mid R_{R\rightarrow L}\mid \,. 
\]

This also forces $S_{LL}S_{RR}^{*}$, \textit{i.e.} $T_{R\rightarrow
L}T_{L\rightarrow R}^{*}$ to be real and 
\begin{eqnarray*}
T_{L\rightarrow R}T_{R\rightarrow L}^{*}+\mid R_{L\rightarrow R}\mid ^2
&=&1\,, \\
T_{R\rightarrow L}T_{L\rightarrow R}^{*}+\mid R_{L\rightarrow R}\mid ^2
&=&1\,.
\end{eqnarray*}

\subsubsection{$H_{\mathcal{T}}=H^\dag$}

Since, by definition, 
\[
H^{\dag} = \left( 
\begin{array}{lr}
H^{*}_{RR} & H^{*}_{LR} \\ 
H^{*}_{RL} & H^{*}_{LL}
\end{array}
\right)\, , 
\]
using the explicit representation of $H_{\mathcal{T}} $, formula (\ref{H_T}%
), we may write 
\begin{equation}
H_{\mathcal{T}} = H^{\dag} + \left( 
\begin{array}{cc}
H_{LL}^{*} - H_{RR}^{*} & 0 \\ 
0 & H_{RR}^{*} - H_{LL}^{*}
\end{array}
\right),  \label{HTHdag}
\end{equation}
which displays transparently the connection between the time reversal
transformed Hamiltonian and the hermitian conjugate of the same Hamiltonian.
They coincide if the diagonal matrix elements $H_{LL} $ and $H_{RR} $ are
identical. In that case the hermiticity of the Hamiltonian forces time
reversal invariance. This is what happens for a local potential, yielding
the current conservation relation (see section~\ref{curr}). For a non-local
potential, in general, $H_{\mathcal{T}}\neq H^\dag$.

If we now assume, instead of time reversal invariance ($H_{\mathcal{T}}=H$),
the condition $H_{\mathcal{T}}=H^{\dag }$, equivalent to the intertwining
relation 
\begin{equation}
H_{\mathcal{T}}\mathcal{T}=\mathcal{T}H=H^{\dag }\mathcal{T}.
\label{T_intertw}
\end{equation}
it is almost immediate to check that 
\begin{equation}
S_{\mathcal{T}}\equiv \mathcal{T}S\mathcal{T}^{-1}=S^{\dag }\,,
\label{S_tau}
\end{equation}
or, in explicit form 
\[
\left( 
\begin{array}{lr}
S_{LL}^{*} & S_{LR}^{*} \\ 
S_{RL}^{*} & S_{RR}^{*}
\end{array}
\right) =\left( 
\begin{array}{lr}
S_{RR}^{*} & S_{LR}^{*} \\ 
S_{RL}^{*} & S_{LL}^{*}
\end{array}
\right) \,, 
\]
leading to the equality of the diagonal matrix elements 
\[
S_{RR}=S_{LL}\,\rightarrow \,T_{L\rightarrow R}=T_{R\rightarrow L}\,. 
\]

\subsubsection{$H_{\mathcal{T}}=H^\dag=H$}

At this stage, we want to point out that if we assume, in addition to $H=H_{%
\mathcal{T}}$, also hermiticity of the Hamiltonian, i.e., $H = H^{\dag} $,
we see from eq.(\ref{HTHdag}) that we must have 
\[
H_{RR} = H^{*}_{RR} = H^{*}_{LL} = H_{LL}\,. 
\]
Thus, the diagonal matrix elements of H are equal and real.

While condition (\ref{T_intertw}) may be met by Hamiltonians that are
neither hermitian, nor $\mathcal{T}$-invariant, for instance those
containing local potentials, it is met a fortiori by Hamiltonians that are
both hermitian and $\mathcal{T}$-invariant. In the latter case, we
simultaneously have $H_{\mathcal{T}}=H\Longrightarrow S_{\mathcal{T}}=S^{-1}$%
, according to Eq. (\ref{S_T_inv}), and $H_{\mathcal{T}}=H^{\dag
}\Longrightarrow S_{\mathcal{T}}=S^{\dag }$, on the basis of Eq. (\ref{S_tau}%
). From these conditions the unitarity of the $S$ matrix follows 
\begin{equation}
S^{\dag }=S^{-1}.  \label{S_dagg}
\end{equation}

In terms of $S$-matrix elements, 
\begin{eqnarray*}
S_{RL}+S_{LR}^{*}\det S &=&0 \\
S_{LR}+S_{RL}^{*}\det S &=&0 \\
S_{LL} &=&S_{RR}^{*}\det S \\
S_{RR} &=&S_{LL}^{*}\det S
\end{eqnarray*}

Summing up, $\mathcal{T}$-invariance and hermiticity of the Hamiltonian lead
to 
\begin{equation}
S^t \mathcal{P} = \mathcal{P} S \makebox[2in]{ and } S^{\dag} = S^{-1}
\end{equation}
or, explicitly, 
\begin{equation}
\det S = - \frac{S_{RL}}{S^{*}_{LR}} = - \frac{ S_{LR}}{S^{*}_{RL}} %
\makebox[1in]{ and } S_{LL} = S_{RR}
\end{equation}
so that 
\begin{equation}
\mid S_{LR} \mid = \mid S_{RL}\mid \makebox[1in]{ and } \mid \det S \mid = 1
\end{equation}

In terms of transmission and reflection coefficients, we then have 
\begin{eqnarray*}
T_{L\rightarrow R} &=& T_{R\rightarrow L} \,, \\
\mid R_{L\rightarrow R}\mid &=& \mid R_{R\rightarrow L}\mid \,.
\end{eqnarray*}

It is worth pointing out that our definition of $\mathcal{T}$ is consistent
with that of ref.~\cite{Mu04}, while ref.~\cite{De03} adopts the following 
\begin{equation}
\mathcal{T}^{\prime}\equiv OK,
\end{equation}
where $K$ is the complex conjugation operator, and $O$ is a unitary operator.

\subsection{$\mathcal{P}\mathcal{T}$ Symmetry}

\label{PTsymm} We may now try to understand the consequences of $\mathcal{P}%
\mathcal{T}$-invariance (but not separately parity or time reversal
invariance).

Any ket in the $|R>$, $|L>$ basis is transformed under $\mathcal{P}\mathcal{T%
}$ according to 
\begin{equation}
\mathcal{P}\mathcal{T}(\rho |R>+\sigma |L>)=\rho ^{*}|R>+\sigma ^{*}|L>
\label{PT_state}
\end{equation}
while, correspondingly, for a bra 
\[
(\rho <R|+\sigma <L|)\mathcal{T}^{-1}\mathcal{P}^{-1}=\rho ^{*}<R|+\sigma
^{*}<L|\quad , 
\]
where $\rho $ and $\sigma $ are complex constants.

The dyadic operators $\Omega_{ij}\equiv |i><j|$, ($i,j=R,L$) are thus
unchanged under $\mathcal{P}\mathcal{T}$ transformations. Since $\mathcal{P}%
\mathcal{T}$ is antilinear, the $\mathcal{P}\mathcal{T}$-transformed
Hamiltonian reads 
\begin{equation}
\begin{array}{ll}
H_{\mathcal{P}\mathcal{T}} = \mathcal{P}\mathcal{T} H \mathcal{T}^{-1}%
\mathcal{P}^{-1} &  \\ 
= H_{RR}^{*} \Omega_{RR} + H_{RL}^{*} \Omega_{RL} + H_{LR}^{*} \Omega_{LR} +
H_{LL}^{*} \Omega_{LL} & .
\end{array}
\label{HTP}
\end{equation}

\subsubsection{$H_{\mathcal{PT}}=H$}

According to Eq. (\ref{HTP}), invariance of the Hamiltonian under $\mathcal{P%
}\mathcal{T}$ leads to the equality 
\begin{equation}
H=\left( 
\begin{array}{lr}
H_{RR} & H_{RL} \\ 
H_{LR} & H_{LL}
\end{array}
\right) =\left( 
\begin{array}{lr}
H_{RR}^{*} & H_{RL}^{*} \\ 
H_{LR}^{*} & H_{LL}^{*}
\end{array}
\right) =H_{P\mathcal{T}}\quad ,  \label{HPTtransf2}
\end{equation}
which shows that the matrix elements have to be real.

From the Schr\"{o}dinger equation (\ref{Schreq}) and the definition of the $%
L-R$ basis, it is immediate to check that, for a local potential, $V(x)$, 
\begin{equation}
H_{RR}=H_{RR}^0+V_{RR}=k^2+\int_{-\infty }^{+\infty
}V(x)dx=H_{LL}^0+V_{LL}=H_{LL}.  \label{maindiag}
\end{equation}

In order for $H_{RR}=H_{LL}$ to be real, it is necessary and sufficient that
the imaginary part of the integral on the \textit{r. h. s}. of Eq. (\ref
{maindiag}) vanishes, \textit{i. e.} the imaginary part of potential $V$ is
an odd function of $x$. As a consequence of the reality of the off-diagonal
matrix elements of $H$, \textit{i. e}. of potential $V$, one readily obtains 
\begin{eqnarray*}
\int_{-\infty }^{+\infty }V_r(x)sin(2kx)dx &=&0 \\
\int_{-\infty }^{+\infty }V_i(x)cos(2kx)dx &=&0.
\end{eqnarray*}
The second relation is automatically fulfilled, since $V_i(x)$ is an odd
function of $x$, while the first relation forces $V_r(x)$ to be an even
function.

It is easy to find how the transition operator defined in eq.(\ref{opevol})
is affected by $\mathcal{P}\mathcal{T}$ invariance (reality) of the
Hamiltonian, $H$. Indeed 
\begin{eqnarray*}
T(t-t_0) &=&\sum_{n=0}^\infty (-i)^nH^n(t-t_0)^n/n! \\
&=&\left( \sum_{n=0}^\infty (i)^nH^n(t-t_0)^n/n!\right) ^{*} \\
&=&\left( T^{-1}(t-t_0)\right) ^{*},
\end{eqnarray*}
and, since 
\begin{equation}
\widetilde{T}^{-1}\left( t-t_0\right) =e^{iH_0t_0}T^{-1}\left( t-t_0\right)
e^{-iH_0t}\;,
\end{equation}
\begin{equation}
\widetilde{T}^{*}\left( t-t_0\right) =e^{-iH_0t}T^{*}\left( t-t_0\right)
e^{iH_0t_0}\;,
\end{equation}
the equality $T^{-1}\left( t-t_0\right) =T^{*}\left( t-t_0\right) $ implies
that 
\[
\lim_{t\rightarrow +\infty ,t_0\rightarrow -\infty }\widetilde{T}^{-1}\left(
t-t_0\right) =\lim_{t\rightarrow +\infty ,t_0\rightarrow -\infty }\widetilde{%
T}^{*}\left( t-t_0\right) \;, 
\]
hence 
\begin{equation}
S^{-1}=S^{*}.  \label{Sinv}
\end{equation}
This yields for the $S$ -matrix elements 
\begin{equation}
\begin{array}{l}
S_{RL}+S_{RL}^{*}\det S=0\;, \\ 
S_{LR}+S_{LR}^{*}\det S=0\;, \\ 
S_{LL}=S_{RR}^{*}\det S\;, \\ 
S_{RR}=S_{LL}^{*}\det S\;.
\end{array}
\label{Sinv_me}
\end{equation}

This imposes that $S_{RL}S_{LR}^{*}=R_{R\rightarrow L}R_{L\rightarrow R}^{*}$
is real, and that 
\begin{equation}
\mid \det S\mid =1\makebox[1in]{ and }\mid S_{LL}\mid =\mid S_{RR}\mid \,,
\label{Det_S_PT}
\end{equation}
where the latter condition corresponds to 
\begin{equation}
\mid T_{L\rightarrow R}\mid =\mid T_{R\rightarrow L}\mid \,.  \label{Eq_T}
\end{equation}

\subsubsection{$H_{\mathcal{PT}}=H$ and $H_{\mathcal{T}}=H^{\dag }$}

\label{H_TH_dag}

Remembering that the time reversed Hamiltonian $H_{\mathcal{T}}$ is, by
definition, such that 
\[
\mathcal{T} H_{\mathcal{T}} = H \mathcal{T} \makebox[1in]{ and } H_{\mathcal{%
T}} \mathcal{T} = \mathcal{T} H 
\]
and assuming, in addition, condition (\ref{T_intertw}), certainly valid for
a local potential and repeated here for clarity's sake, ${\mathcal{T}} H =
H^{\dag} \mathcal{T} $, forces the diagonal elements of the Hamiltonian
matrix to be equal, $H_{LL} = H_{RR} $ ; we then see that for a $\mathcal{P}%
\mathcal{T}$-invariant Hamiltonian the following relations hold 
\[
\mathcal{P} H^{\dag} = H \mathcal{P} \makebox[1in]{ or } H^\dag \mathcal{P}
= \mathcal{P} H 
\]

On the conditions given above, 
\[
\mathcal{P}H\mathcal{P}^{-1}=H^{\dag },
\]
\textit{i.e., }$H$ is pseudo-hermitian~\cite{Mo02} with respect to $\mathcal{%
P}$.

As a consequence, the transition operator behaves as follows 
\begin{eqnarray*}
\mathcal{P}T^{\dag }(t-t_0) &=&\sum_{n=0}^\infty (i)^n\mathcal{P}H{^{\dag }}%
^n(t-t_0)^n/n! \\
&=&\sum_{n=0}^\infty (i)^nH^n(t-t_0)^n\mathcal{P}/n! \\
&=&T^{-1}(t-t_0)\mathcal{P}.
\end{eqnarray*}
Hence, recalling Eq. (\ref{TrOp}), 
\begin{eqnarray}
\mathcal{P}\widetilde{T}^{\dagger }\left( t-t_0\right) &=&\mathcal{P}%
e^{iH_0t_0}e^{iH^{\dagger }\left( t-t_0\right) }e^{-iH_0t} \\
&=&\mathcal{P}e^{iH_0t_0}e^{i\mathcal{P}H\mathcal{P}^{-1}\left( t-t_0\right)
}e^{-iH_0t} \\
&=&\mathcal{P}^2e^{iH_0t_0}e^{iH\left( t-t_0\right) }e^{-iH_0t}\mathcal{P}%
^{-1} \\
&=&e^{iH_0t_0}e^{iH\left( t-t_0\right) }e^{-iH_0t}\mathcal{P} \\
&=&\widetilde{T}^{-1}\left( t-t_0\right) \mathcal{P}
\end{eqnarray}
and, in the $t_0\rightarrow -\infty $, $t\rightarrow +\infty $ limits, 
\begin{equation}
\mathcal{P}S^{\dag }=S^{-1}\mathcal{P}
\end{equation}

Thus, under both $\mathcal{P}\mathcal{T}$-invariance of the Hamiltonian and
the condition that $H_{\mathcal{T}}=H^{\dag }$, so that the diagonal matrix
elements of the Hamiltonian are real and equal, $H_{LL}=H_{RR}$, we have,
since $S^{-1}=S^{*}$, 
\begin{equation}
\mathcal{P}S^{\dag }=S^{*}\mathcal{P}\makebox[1in]{ or }S^t\mathcal{P}=%
\mathcal{P}S\,.
\end{equation}
In turn, this leads to the equality of the diagonal elements of the S matrix 
\begin{equation}
S_{RR}=S_{LL}\Longrightarrow T_{L\rightarrow R}=T_{R\rightarrow L}\,.
\label{diag_el}
\end{equation}
Thus 
\begin{eqnarray*}
S_{RL}+S_{RL}^{*}\det S &=&0 \\
S_{LR}+S_{LR}^{*}\det S &=&0 \\
S_{LL} &=&S_{RR}=S_{LL}^{*}\det S=S_{RR}^{*}\det S\,, \\
&&
\end{eqnarray*}
which leads to 
\[
\det S=\frac{S_{RR}}{S_{RR}^{*}}=\frac{S_{LL}}{S_{LL}^{*}}=-\frac{S_{RL}}{%
S_{RL}^{*}}=-\frac{S_{LR}}{S_{LR}^{*}}\,, 
\]
or, in terms of transmission and reflection coefficients, 
\begin{eqnarray*}
\mid \det S\mid &=\left| T_{L\rightarrow R}/T_{L\rightarrow R}^{*}\right|
=&1, \\
S_{LR}S{_{RL}^{*}} &=&S_{RL}S_{LR}^{*}\,\rightarrow \,R_{L\rightarrow
R}R_{R\rightarrow L}^{*}=R_{R\rightarrow L}R_{L\rightarrow R}^{*}\;, \\
\,R_{L\rightarrow R}+\,R_{L\rightarrow R}^{*}T_{L\rightarrow
R}/T_{L\rightarrow R}^{*} &=&0\;, \\
\,R_{R\rightarrow L}+\,R_{R\rightarrow L}^{*}T_{R\rightarrow
L}/T_{R\rightarrow L}^{*} &=&0\;.
\end{eqnarray*}

It is worthwhile to stress again that the equality (\ref{diag_el}) of the
two transmission coefficients is not a consequence of $\mathcal{PT}$
symmetry, which yields only the equality of their moduli, but of the
additional intertwining condition (\ref{T_intertw}), valid for any local
potential. Again, this point is discussed in Ref.~\cite{Mu04}.

\subsubsection{Exact Asymptotic $\mathcal{P}\mathcal{T}$ Symmetry}

$\label{Exact_PT}$

In this sub-section, we consider Hamiltonians with exact $\mathcal{P}%
\mathcal{T}$ symmetry, \textit{i. e.} $\mathcal{P}\mathcal{T}$-invariant
Hamiltonians whose eigenstates are also eigenstates of $\mathcal{P}\mathcal{T%
}$.

It has been recently proved~\cite{Mo03} that a Hamiltonian with exact $%
\mathcal{P}\mathcal{T}$ symmetry is unitarily equivalent to a Hamiltonian
that is hermitian with respect to a suitably defined inner product.

Let us now investigate some consequences of exact $\mathcal{PT}$ symmetry on
the $S$ matrix. To this aim, it is convenient to introduce the
transformation under $\mathcal{P}\mathcal{T}$ of a generic wave function, $%
\Psi \left( x\right) $%
\begin{equation}
\mathcal{P}\mathcal{T}\Psi \left( x\right) \equiv \Psi _{\mathcal{P}\mathcal{%
T}}\left( x\right) =\Psi ^{*}\left( -x\right)
\end{equation}
and the condition of exact $\mathcal{PT}$ symmetry 
\begin{equation}
\Psi _{\mathcal{P}\mathcal{T}}\left( x\right) =\Psi ^{*}\left( -x\right)
=e^{i\theta }\Psi \left( x\right) \;,  \label{PT_ex}
\end{equation}
where $\theta $ is a real number, because $\left( \mathcal{P}\mathcal{T}%
\right) ^2=1$.

Let us apply Eq. (\ref{PT_ex}) to the asymptotic wave functions defined in
section 2, 
\begin{equation}
\Psi _{\mathcal{P}\mathcal{T}}\left( \pm \infty \right) =\Psi ^{*}\left( \mp
\infty \right) =e^{i\theta }\Psi \left( \pm \infty \right) \;,
\end{equation}
and, in terms of their amplitudes, $A_{\pm },B_{\pm },\widetilde{A}_{\pm }$
and$\widetilde{\text{ }B}_{\pm }$, 
\begin{eqnarray}
A_{\pm }^{*}e^{ikx}+B_{\pm }^{*}e^{-ikx} &=&e^{i\theta }\left( A_{\mp
}e^{ikx}+B_{\mp }e^{-ikx}\right) \;, \\
\widetilde{A}_{\pm }^{*}e^{ikx}+\widetilde{B}_{\pm }^{*}e^{-ikx} &=&e^{i%
\widetilde{\theta }}\left( \widetilde{A}_{\mp }e^{ikx}+\widetilde{B}_{\mp
}e^{-ikx}\right) \;.
\end{eqnarray}

Hence, in particular 
\begin{equation}
A_{+}^{*}=e^{i\theta }A_{-}\;,\;\widetilde{B}_{+}^{*}=e^{i\widetilde{\theta }%
}\widetilde{B}_{-}\;,  \label{PT_ex_0}
\end{equation}
and 
\begin{equation}
B_{-}=e^{-i\theta }B_{+}^{*}=0\;,\;\widetilde{A}_{+}^{*}=e^{i\widetilde{%
\theta }}\widetilde{A}_{-}=0\;.  \label{PT_ex_2}
\end{equation}

The latter equalities come from the boundary conditions for an incident
progressive wave $\left( B_{+}=0\right) $ and an incident regressive wave $%
\left( \widetilde{A}_{-}=0\right) $, respectively. The $S$ matrix elements
are thus written as 
\begin{equation}
\begin{array}{l}
S_{RR}=T_{L\rightarrow R}=\frac{A_{+}}{A_{-}}=e^{-^{i\theta }}\frac{A_{-}^{*}%
}{A_{-}}=e^{-i\left( \theta +2\alpha _{-}\right) },\;\left( A_{-}=\left|
A_{-}\right| e^{i\alpha _{-}}\right) \\ 
S_{LR}=R_{L\rightarrow R}=\frac{B_{-}}{A_{-}}=0\;, \\ 
S_{LL}=T_{R\rightarrow L}=\frac{\widetilde{B}_{-}}{\widetilde{B}_{+}}=e^{-i%
\widetilde{\theta }}\frac{\widetilde{B}_{+}^{*}}{\widetilde{B}_{+}}%
=e^{-i\left( \widetilde{\theta }+2\widetilde{\beta _{+}}\right) }\;,\;\left( 
\widetilde{B}_{+}=\left| \widetilde{B}_{+}\right| e^{i\widetilde{\beta _{+}}%
}\right) \\ 
S_{RL}=R_{R\rightarrow L}=\frac{\widetilde{A}_{+}}{\widetilde{B}_{+}}=0.
\end{array}
\label{PT_ex_1}
\end{equation}

The reflection coefficients are thus zero and the transmission coefficients
have unit modulus, in keeping with the more general condition $\left| \det
S\right| =1$, Eq. (\ref{Det_S_PT}), imposed by the $\mathcal{PT}$ symmetry
of the Hamiltonian. In this case, the $S$ matrix is unitary, too, since $%
S^{-1}=S^{*}=\left( S^{*}\right) ^t=S^{\dagger }$.

Conversely, it is not difficult to show that conditions (\ref{PT_ex_1}) on
the $S$ matrix elements are sufficient to ensure that the corresponding
asymptotic wave functions are eigenstates of $\mathcal{PT}$.

Examples of $\mathcal{PT}$-symmetric reflectionless potentials are discussed
in section \ref{refless}.

\section{Probability Current and Density for Linear and Antilinear
Transformations}

\label{curr}

In the present section, we consider the time-dependent Schr\"{o}dinger
equation (\ref{Schreqt}) with a local potential, $V(x)$, and its solution, $%
\psi (x,t)$. In this case, $\mathcal{T}{}$ invariance is equivalent to
hermiticity and $V$ is real.

We introduce a linear transformation, $U_L,$ and a corresponding antilinear
transformation, $U_A$, commuting with the kinetic energy operator, $%
p^2=-\partial ^2/\partial x^2$, and apply them to the Schr\"{o}dinger
equation (\ref{Schreqt}) with a local potential (\ref{local}): 
\begin{eqnarray*}
\left( -\frac{\partial ^2}{\partial x^2}+V_{U_L}\left( x\right) \right) \psi
_{U_L}\left( x,t\right)  &=&i\frac \partial {\partial t}\psi _{U_L}\left(
x,t\right) \,; \\
\left( -\frac{\partial ^2}{\partial x^2}+V_{U_A}\left( x\right) \right) \psi
_{U_A}\left( x,t\right)  &=&-i\frac \partial {\partial t}\psi _{U_A}\left(
x,t\right) \,.
\end{eqnarray*}
where $\Psi _U\equiv U\Psi $ and 
\[
V_U(x)\equiv UV(x)U^{-1}\,,
\]
or, equivalently 
\[
V_U(x)U=UV\left( x\right) \;,
\]
namely, if $V$ is invariant under $U$, it commutes with $U$.

Now, we multiply the equation satisfied by $\psi _{U_A}$ by $\psi $ and the
initial Schr\H{o}dinger equation by $\psi _{U_A}$ and subtract them side by
side, obtaining: 
\begin{equation}
-\left( \psi _{U_A}\frac{\partial ^2}{\partial x^2}\psi -\psi \frac{\partial
^2}{\partial x^2}\psi _{U_A}\right) +\left( V-V_{U_A}\right) \psi _{U_A}\psi
=i\frac \partial {\partial t}\left( \psi _{U_A}\psi \right)\,.  \label{Ceq}
\end{equation}

Now we introduce the density of probability current: 
\[
j_{U_A}(x,t)\equiv -i\left( \left( \frac \partial {\partial x}\psi \left(
x,t\right) \right) \psi _{U_A}\left( x,t\right) -\psi \left( x,t\right)
\left( \frac \partial {\partial x}\psi _{U_A}\left( x,t\right) \right)
\right) \,, 
\]
and the density of probability: 
\[
\rho _{U_A}\left( x,t\right) \equiv \psi \left( x,t\right) \psi _{U_A}\left(
x,t\right) \,. 
\]

Therefore, Eq. (\ref{Ceq}) can be rewritten as: 
\begin{equation}
\frac \partial {\partial x}j_{U_A}+\frac \partial {\partial t}\rho
_{U_A}=-i\left( V-V_{U_A}\right) \rho _{U_A}\,.  \label{Ceq1}
\end{equation}

If the potential is invariant under $U_A$ the right-hand side of Eq. (\ref
{Ceq1}) is zero and we obtain a continuity equation, which, for stationary
waves, yields a constant current.

Analogously, for a linear transformation, we first have to consider the
complex conjugate of the equation satisfied by $\psi _{U_L}$ and repeat the
same steps as before, by replacing everywhere $\psi _{U_A}$ by $\psi
_{U_L}^{*}$ and $V_{U_A}$ by $V_{U_L}^{*}:$%
\begin{equation}
\frac \partial {\partial x}j_{U_L}+\frac \partial {\partial t}\rho
_{U_L}=-i\left( V-V_{U_L}^{*}\right) \rho _{U_L}\,.  \label{Ceq2}
\end{equation}

For real $V$, similar considerations are valid and Eq. (\ref{Ceq2}) is
reduced to a continuity equation.

The formalism outlined above does not apply to non-local potentials. An
attempt to extend flux conservation to the latter case is described in Ref.%
\cite{Ao83}.

\subsection{The $\mathcal{PT}$-Symmetric Case}

In the case of $\mathcal{PT}$-symmetric potentials, the definitions of $\rho 
$ and $j$ satisfying a continuity equation are easily obtained by writing
the time-dependent Schr\H{o}dinger equations satisfied by $\psi _{\mathcal{P}%
}\left( x,t\right) $ and $\psi _{\mathcal{T}}\left( x,t\right) $: 
\begin{eqnarray*}
\left( -\frac{\partial ^2}{\partial x^2}+V\left( -x\right) \right) \psi _{%
\mathcal{P}}\left( x,t\right) &=&i\frac \partial {\partial t}\psi _{\mathcal{%
P}}\left( x,t\right) \;, \\
\left( -\frac{\partial ^2}{\partial x^2}+V^{*}\left( x\right) \right) \psi _{%
\mathcal{T}}\left( x,t\right) &=&-i\frac \partial {\partial t}\psi _{%
\mathcal{T}}\left( x,t\right) \;,
\end{eqnarray*}
where $\psi _{\mathcal{P}}\left( x,t\right) =\psi \left( -x,t\right) $ and $%
\psi _{\mathcal{T}}\left( x,t\right) =\psi ^{*}\left( x,t\right) $, so that
the equations given above are equivalent to the following 
\begin{eqnarray}
\left( -\frac{\partial ^2}{\partial x^2}+V\left( -x\right) \right) \psi
\left( -x,t\right) &=&i\frac \partial {\partial t}\psi \left( -x,t\right) \;,
\\
\left( -\frac{\partial ^2}{\partial x^2}+V^{*}\left( x\right) \right) \psi
^{*}\left( x,t\right) &=&-i\frac \partial {\partial t}\psi ^{*}\left(
x,t\right) \;.
\end{eqnarray}

By multiplying side by side the first equation by $\psi ^{*}\left(
x,t\right) $ and the second equation by $\psi \left( -x,t\right) $, and
subtracting the two equations side by side, we obtain 
\begin{eqnarray}
&&-\psi ^{*}\left( x,t\right) \frac{\partial ^2}{\partial x^2}\psi \left(
-x,t\right) +\psi \left( -x,t\right) \frac{\partial ^2}{\partial x^2}\psi
^{*}\left( x,t\right) +\left( V\left( -x\right) -V^{*}\left( x\right)
\right) \psi ^{*}\left( x,t\right) \psi \left( -x,t\right)  \nonumber \\
&=&i\psi ^{*}\left( x,t\right) \frac \partial {\partial t}\psi \left(
-x,t\right) +i\psi \left( -x,t\right) \frac \partial {\partial t}\psi
^{*}\left( x,t\right) \;.  \nonumber
\end{eqnarray}

${\mathcal{PT}}$ invariance implies that $V\left( -x\right) =V^{*}\left(
x\right) $, so that the equation above can be reduced to a continuity
equation 
\[
\frac \partial {\partial x}j\left( x,t\right) +\frac \partial {\partial
t}\rho \left( x,t\right) =0\,, 
\]
with 
\begin{equation}
\rho \left( x,t\right) =\psi ^{*}\left( x,t\right) \psi \left( -x,t\right)
\,,  \label{PT_rho}
\end{equation}
\begin{equation}
j\left( x,t\right) =\frac 1i[\psi ^{*}\left( x,t\right) \frac \partial
{\partial x}\psi \left( -x,t\right) -\psi \left( -x,t\right) \frac \partial
{\partial x}\psi ^{*}\left( x,t\right) ]\,\;,  \label{PTj}
\end{equation}
which is consistent with the definition of $j$ given in Ref.~\cite{BQZ01}.
The following relation holds: 
\[
j^{*}(-x,t)=-j\left( x,t\right) . 
\]
It is worthwhile to point out that $j$ is identically zero when $\psi
^{*}\left( x\right) =e^{i\alpha }\psi \left( -x\right) ,$ where $\alpha $ is
a real constant, i. e. $\psi $ is an eigenstate of $\mathcal{PT}$. If $\psi $
is a stationary wave, $\psi \left( x,t\right) =e^{-iEt}\psi \left( x\right)
, $ $j$ depends only on $x$ and current (\ref{PTj}) is constant in the whole
space. The asymptotic solutions can be taken in the form of stationary waves
and we obtain, remembering that, for a wave travelling from left to right,
Eq.(\ref{Progwave}), we have $B_{+}=0$: 
\begin{eqnarray*}
j\left( +\infty \right) &=&\frac 1i\left\{ 
\begin{array}{c}
A_{+}^{*}e^{-ikx}(-ik)\left[ A_{-}e^{-ikx}-B_{-}e^{+ikx}\right] \\ 
-\left[ A_{-}e^{-ikx}+B_{-}e^{+ikx}\right] \left( -ik\right)
A_{+}^{*}e^{-ikx}
\end{array}
\right\} \\
&=&2kA_{+}^{*}B_{-}\,,
\end{eqnarray*}
and, in the same way 
\[
j\left( -\infty \right) =-2kA_{+}B_{-}^{*}\,. 
\]

As expected, $j\left( -\infty \right) =-j^{*}\left( +\infty \right) $. Note
that, for the reflectionless potentials discussed in sub-section \ref
{Exact_PT}, $B_{-}=0$ . In this case, $j\left( -\infty \right) =j\left(
+\infty \right) =0$.

Assuming now $B_{-}\neq 0$, conservation of current implies that we also
have $j\left( -\infty \right) =j\left( +\infty \right) $; hence, the current
is purely imaginary and we can write: 
\[
A_{+}B_{-}^{*}+A_{+}^{*}B_{-}=0\,, 
\]
or, dividing both sides of the above equation by $A_{-}A_{-}^{*}$ and
remembering the definitions of $T_{L\rightarrow R}$ and $R_{L\rightarrow R}$%
: 
\[
R_{L\rightarrow R}T_{L\rightarrow R}^{*}+R_{L\rightarrow
R}^{*}T_{L\rightarrow R}=0\,, 
\]
a result already obtained in sub-section \ref{H_TH_dag}, or, equivalently: 
\[
\frac{R_{L\rightarrow R}}{R_{L\rightarrow R}^{*}}+\frac{T_{L\rightarrow R}}{%
T_{L\rightarrow R}^{*}}=0\,, 
\]
which means that the phase $\varphi _r$ of the reflection coefficient and $%
\varphi _t$ of the transmission coefficient are related by: 
\[
\varphi _r=\varphi _t+\frac \pi 2+n\pi , 
\]
where $n$ is an integer. This relation is trivially checked for square well 
\cite{Me98}, hyperbolic Scarf~\cite{Le01} and generalized P\H{o}schl-Teller~%
\cite{Le02} potentials.

\section{Explicit Relations for Some Potentials}

\label{examples}

For the sake of example, the formalism worked out in the previous sections
is here applied in detail to a few $\mathcal{PT}$-symmetric solvable
potentials. For those already considered in Ref. (\cite{De03}), having an
imaginary part made of an odd combination of Dirac delta functions, $\Im
V(x)=\lambda \left( \delta \left( x+\frac{X_0}2\right) -\delta \left( x-%
\frac{X_0}2\right) \right) $, it is easy to check that all the relations of
subsection \ref{PTsymm} are valid.

In the presentation of the examples, we make a selection dictated by
solvability implemented by different methods.

\subsection{Complex $\mathcal{PT}$-Symmetric Square Well}

\label{sec-square well}

Let 
\begin{equation}
V=0,\qquad x<-b\;,\quad x>+b  \label{Sq_well}
\end{equation}
\[
V=-V_0+iV_1\,,\qquad -b\leq x\leq 0\,, 
\]
\[
V=-V_0-iV_1\,,\qquad 0\leq x\leq +b\,, 
\]
Here, $V_0$ and $V_0$ are real parameters and $V_0\geq 0$.

It is worthwhile to mention that discrete states in a $\mathcal{PT}$%
-symmetric square well with $V_0=0$ were already studied in Ref.\cite{Zn01}.
It was shown, in particular, that the square well possesses a real discrete
spectrum on condition that the coefficient, $V_1$, of the imaginary part is
smaller than a certain critical value. On this condition, the eigenfunctions
of $H$ are also eigenfunctions of $\mathcal{PT}$, i. e. the model has an
exact $\mathcal{PT}$ symmetry. In this case, it is possible to construct a
hermitian Hamiltonian unitarily equivalent to the $\mathcal{PT}$-symmetric
square well in a Hilbert space endowed with a properly defined scalar
product, as shown in full detail in Ref.~\cite{MB04}. The analysis has been
extended to the continuum of scattering states in Ref.\cite{Mo05}.

Introducing now 
\[
k^2 = E 
\]
and 
\[
\alpha_{0}^{2} = E + V_{0} - i V_{1} 
\]
\[
\alpha_{1}^{2} = E +V_{0} + i V_{1}\,, 
\]
implying 
\[
\alpha_{0,1} = \alpha \exp{(\mp i\varphi)}\,, 
\]
\[
\alpha^*_{0} = \alpha_{1}, \hspace{1cm} \alpha^*_{1} = \alpha_{0}\,, 
\]
with 
\begin{eqnarray*}
\alpha^2 & = & \sqrt{\left( E+V_0 \right)^2 + V_1^2} \\
\varphi & = & \frac{1}{2} \arctan \left( \frac{V_1}{E +V_0} \right)
\end{eqnarray*}

The two linearly independent solutions (Eqs. (\ref{Progwave}-\ref{XY}) can
both be written in the general form 
\[
\Psi (x)=\left\{ 
\begin{array}{lcr}
Ae^{ikx}+Be^{-ikx} &  & \mbox{$x \leq {- b}$} \\ 
Ce^{i\alpha _0x}+De^{-i\alpha _0x} &  & \mbox{ $-b \leq x \leq 0 $} \\ 
Ee^{i\alpha _1x}+Fe^{-i\alpha _1x} &  & \mbox{$0 \leq x \leq b $} \\ 
Ge^{ikx}+He^{-ikx} &  & \mbox{$b \leq x$}
\end{array}
\right. 
\]
with suitable specification of parameters $A,...,H$.

The $\mathcal{PT}$-transformed wavefunction reads 
\[
\Psi _{\mathcal{PT}}(x)=\left\{ 
\begin{array}{lcr}
G^{*}e^{ikx}+H^{*}e^{-ikx} &  & \mbox{$x \leq {- b}$} \\ 
E^{*}e^{i\alpha _0x}+F^{*}e^{-i\alpha _0x} &  & \mbox{ $-b \leq x \leq 0 $}
\\ 
C^{*}e^{i\alpha _1x}+D^{*}e^{-i\alpha _1x} &  & \mbox{$0 \leq x \leq b $} \\ 
A^{*}e^{ikx}+B^{*}e^{-ikx} &  & \mbox{$b \leq x$}
\end{array}
\right. 
\]

$\mathcal{PT}$-symmetric wave functions would thus meet the conditions $%
A=G^{*}$, $B=H^{*}$, $C=E^{*}$ and $D=F^{*}$. In the treatment of scattering
states for $V_0=0$, Ref.\cite{Mo05} considers two linearly independent $%
\mathcal{PT}$-symmetric wave functions, in keeping with the double
degeneracy and reality of the eigenvalues $E=k^2$. We stress, however, that
these $\mathcal{PT}$-symmetric eigenfunctions are not the $\Psi _1$ and $%
\Psi _2$ functions (see Eqs. (\ref{Psi_1}-\ref{Psi_2}) studied in section 
\ref{Exact_PT}.

We then have three pairs of continuity equations at $x= -b$, $x = 0 $ and $x
= b$, respectively : 
\[
\left( 
\begin{array}{lr}
e^{-ikb} & e^{ikb} \\ 
e^{-ikb} & - e^{ikb}
\end{array}
\right) \left( 
\begin{array}{l}
A \\ 
B
\end{array}
\right) = \left( 
\begin{array}{lr}
e^{-i{\alpha_0} b} & e^{i\alpha_0 b} \\ 
\frac{\alpha_0}{k} e^{-i\alpha_0 b} & -\frac{\alpha_0}{k} e^{i\alpha_0 b}
\end{array}
\right) \left( 
\begin{array}{l}
C \\ 
D
\end{array}
\right) \,, 
\]

\[
\left( 
\begin{array}{lr}
1 & 1 \\ 
1 & - 1
\end{array}
\right) \mbox{} \left( 
\begin{array}{l}
C \\ 
D
\end{array}
\right) = \left( 
\begin{array}{lr}
1 & 1 \\ 
\frac{\alpha_1}{\alpha_0} & -\frac{\alpha_1}{\alpha_0}
\end{array}
\right) \left( 
\begin{array}{l}
E \\ 
F
\end{array}
\right) \,, 
\]
implying 
\[
\mid C\mid ^2 + \mid D\mid ^2 = \mid E \mid ^2 + \mid F\mid ^2 \,, 
\]

\[
\left( 
\begin{array}{lr}
e^{i{\alpha_1}b} & e^{-i{\alpha_1}b} \\ 
e^{i{\alpha_1}b} & - e^{-i{\alpha_1}b}
\end{array}
\right) \mbox{}\left( 
\begin{array}{l}
E \\ 
F
\end{array}
\right) = \left( 
\begin{array}{lr}
e^{ik b} & e^{-ik b} \\ 
\frac{k}{\alpha_1} e^{ikb} & -\frac{k}{\alpha_1} e^{-ik b}
\end{array}
\right) \left( 
\begin{array}{l}
G \\ 
H
\end{array}
\right)\,. 
\]

They allow to re-express the coefficients of the wave function at $x=-\infty$
in terms of those at $x=+\infty$ 
\[
\left( 
\begin{array}{l}
A \\ 
B
\end{array}
\right) = \left( 
\begin{array}{lr}
M_{RR} & M_{RL} \\ 
M_{LR} & M_{LL}
\end{array}
\right) \left ( 
\begin{array}{l}
G \\ 
H
\end{array}
\right ) \,, 
\]
with 
\[
M_{i,j} = M_{i,j} ({\alpha_0},{\alpha_1}, b, k)\qquad (i,j=R,L) 
\]

The various matrix elements are given by the following expressions 
\begin{eqnarray*}
M_{RR} = \frac{1}{8} e^{2ikb} \left\{ f({\alpha_0},{\alpha_1},b,k)+ f({%
\alpha_0},-{\alpha_1},b,k) \right.  \nonumber \\
\left. + f(-{\alpha_0},{\alpha_1},b,k)+f(-{\alpha_0},-{\alpha_1},b,k)
\right\}
\end{eqnarray*}

\begin{eqnarray*}
M_{LR} = \frac{1}{8} \left\{ g({\alpha_0},{\alpha_1},b,k)+g({\alpha_0},-{%
\alpha_1},b,k) \right.  \nonumber \\
\left. + g(-{\alpha_0},{\alpha_1},b,k)+g(-{\alpha_0},-{\alpha_1},b,k)
\right\}
\end{eqnarray*}

\begin{eqnarray*}
M_{RL} = \frac{1}{8} \left\{ h({\alpha_0},{\alpha_1},b,k)+h({\alpha_0},-{%
\alpha_1},b,k) \right.  \nonumber \\
\left. + h(-{\alpha_0},{\alpha_1},b,k)+h(-{\alpha_0},-{\alpha_1},b,k)
\right\}
\end{eqnarray*}
and 
\begin{eqnarray*}
M_{LL} = \frac{1}{8} e^{-2ikb} \left\{ m({\alpha_0},{\alpha_1},b,k) + m({%
\alpha_0},-{\alpha_1},b,k) \right.  \nonumber \\
\left. + m(-{\alpha_0},{\alpha_1},b,k)+m(-{\alpha_0},-{\alpha_1},b,k)
\right\}
\end{eqnarray*}

with 
\[
f({\alpha_0},{\alpha_1},b,k) = (1 + \frac{\alpha_1}{\alpha_0})(1 + \frac{%
\alpha_0}{k}) (1 + \frac{k}{\alpha_1}) \mbox{} e^{-i({\alpha_0} + {\alpha_1}%
)b} 
\]
and 
\[
g({\alpha_0},{\alpha_1},b,k) = (1 + \frac{\alpha_1}{\alpha_0})(1 - \frac{%
\alpha_0}{k}) (1 + \frac{k}{\alpha_1}) \mbox{} e^{-i({\alpha_0} + {\alpha_1}%
)b} 
\]
\[
h({\alpha_0},{\alpha_1},b,k) = (1 + \frac{\alpha_1}{\alpha_0})(1 + \frac{%
\alpha_0}{k}) (1 - \frac{k}{\alpha_1}) \mbox{} e^{-i({\alpha_0} + {\alpha_1}%
)b} 
\]
and 
\[
m({\alpha_0},{\alpha_1},b,k) = (1 + \frac{\alpha_1}{\alpha_0})(1 - \frac{%
\alpha_0}{k}) (1 - \frac{k}{\alpha_1}) \mbox{} e^{-i({\alpha_0} + {\alpha_1}%
)b} 
\]

The obvious symmetries relating these auxiliary functions stem out 
\[
m({\alpha_0},{\alpha_1},b,k) = f(-{\alpha_0},-{\alpha_1},-b,k) 
\]
\[
h({\alpha_0},{\alpha_1},b,k) = g(-{\alpha_0},-{\alpha_1},-b,k) 
\]
From these definitions, we see that 
\[
f({\alpha _0},{\alpha _1},b,k)=f({\alpha _1},{\alpha _0},b,k)=f({-\alpha _0},%
{-\alpha _1},-b,-k)=f^{*}({\alpha _1},{\alpha _0},-b,k) 
\]
\begin{eqnarray*}
h({\alpha _0},{\alpha _1},b,k) &=&-h({\alpha _1},{\alpha _0},b,-k)=-h({%
-\alpha _1},{-\alpha _0},-b,k) \\
&=&h({-\alpha _0},{-\alpha _1},-b,-k)=-h^{*}({\alpha _0},{\alpha _1},-b,-k)
\end{eqnarray*}

A series of relations concerning the $M$-matrix elements follow from the
above relations: some of which are listed below, not pretending to
completeness (note that $M_{RR}$ and $M_{LL}$ are invariant under the
exchange of $\alpha_0$ and $\alpha_1$) : 
\begin{eqnarray*}
M_{RR} ({\alpha_0},{\alpha_1},b,k) = M_{RR} ({\alpha_1},{\alpha_0},b,k) =
M_{RR} (-{\alpha_0},{-\alpha_1},-b,-k) \\
= M^*_{RR} ({\alpha_0},{\alpha_1},b,- k)
\end{eqnarray*}
and correspondingly for $M_{LL}$ with, in addition, 
\[
M_{LL} ({\alpha_0},{\alpha_1}, b, k) = M_{RR} ({\alpha_0},{\alpha_1},b, - k)
= M_{RR} (-{\alpha_0},{-\alpha_1},-b, k) \,, 
\]
while 
\begin{eqnarray*}
M_{RL} ({\alpha_0},{\alpha_1},b,k) = - M_{RL} ({\alpha_1},{\alpha_0},b,-k) =
M_{RL} (-{\alpha_0},{-\alpha_1},-b,-k) \\
=- M^*_{RL} ({\alpha_0},{\alpha_1},- b,- k)
\end{eqnarray*}
and correspondingly for $M_{LR}$ with, in addition, 
\begin{eqnarray*}
M_{LR} ({\alpha_0},{\alpha_1}, b, k) = M_{RL} (-{\alpha_0},-{\alpha_1},-b,
k) = - M_{RL} ({\alpha_1},{\alpha_0},b, k) \\
= - M^*_{RL} ({\alpha_0},{\alpha_1}, - b, k)
\end{eqnarray*}

We can then give more explicit expressions for the diagonal matrix elements
: 
\begin{eqnarray*}
M_{RR} & =& e^{2ikb}\cdot \left\{ \cos^2\varphi \cos (2\alpha b \cos \varphi
)+\sin^2\varphi\cosh (2\alpha b \sin\varphi)\right. -i\frac{k^2-\alpha^2} {%
2k\alpha} \cdot \\
& &\left.\hspace{1cm}\cdot \sin{\varphi}\sinh(2\alpha b\sin\varphi) -i\frac{%
k^2+\alpha^2}{2k\alpha}\cos{\varphi} \sin(2\alpha b\cos \varphi)\right\}
\end{eqnarray*}
and 
\begin{eqnarray*}
M_{LL} & =& e^{-2ikb}\cdot \left\{ \cos^2 \varphi \cos (2\alpha b\cos\varphi
)+\sin^2\varphi \cosh(2\alpha b \sin\varphi) +i\frac{k^2-\alpha^2} {2k\alpha}%
\cdot \right. \\
& & \left.\hspace{1cm} \cdot \sin{\varphi}\sinh(2\alpha b\sin\varphi) +i%
\frac{k^2+\alpha^2}{2k\alpha}\cos{\varphi}\sin(2\alpha b\cos\varphi)\right\}
\,,
\end{eqnarray*}
whereas the non diagonal matrix elements read 
\begin{eqnarray*}
M_{RL}&=& i\left\{\sin\varphi\cos\varphi[\cos(2\alpha
b\cos\varphi)-\cosh(2\alpha b\sin\varphi)]+\frac{k^2-\alpha^2}{2k\alpha}%
\right. \cdot \\
& & \cdot\left.\cos\varphi\sin(2\alpha b\cos\varphi)+\frac{k^2+\alpha^2}{%
2k\alpha}\sin\varphi\sinh(2\alpha b\sin\varphi)\right\}
\end{eqnarray*}
and 
\begin{eqnarray*}
M_{LR}&=& i\left\{ \sin\varphi\cos\varphi [\cos(2\alpha
b\cos\varphi)-\cosh(2\alpha b\sin\varphi)]-\frac{k^2-\alpha^2}{2k\alpha}%
\cdot \right. \\
& & \cdot \left.\cos\varphi \sin(2\alpha b\cos\varphi)-\frac{k^2+\alpha^2}{%
2k\alpha}\sin\varphi\sinh(2\alpha b\sin\varphi)\right\}\,,
\end{eqnarray*}
where we have used the modulus $\alpha$ and the phase ${\varphi}$ introduced
earlier.

Transmission and reflection coefficients are easily expressed in terms of
the $M$-matrix elements, on the basis of Eq. ( \ref{M_mat}).

Assuming an incident wave coming from the left, i.e., from $x=-\infty$, the
reflection and transmission amplitudes are obtained by assuming $H = 0$ 
\[
R_{L\rightarrow R} = \frac{M_{LR}}{M_{RR}} 
\]
and 
\[
T_{L\rightarrow R} = \frac{1}{M_{RR}}\,, 
\]
which implies that 
\[
\mid R_{L\rightarrow R} \mid ^2 + \mid T_{L\rightarrow R} \mid ^2 = \frac{1
+ \mid M_{LR} \mid ^2}{\mid M_{RR} \mid ^2} 
\]
and 
\[
R_{L\rightarrow R} \hspace{.1cm} T^*_{L\rightarrow R} + R^*_{L\rightarrow R} 
\hspace{.1cm} T_{L\rightarrow R} = 2 \hspace{.1cm} Re(M_{LR})\,. 
\]

Correspondingly, we have for an incident wave coming from $x=+\infty$ 
\[
R_{R\rightarrow L} = - \frac{M_{RL}}{M_{RR}} 
\]
and 
\[
T_{R\rightarrow L} = \frac{\det\hspace{.1cm} M \hspace{.1cm}}{M_{RR}} = 
\frac{1}{M_{RR}}\,, 
\]
which implies that 
\[
\mid R_{R\rightarrow L} \mid ^2 + \mid T_{R\rightarrow L} \mid ^2 = \frac{1
+ \mid M_{RL} \mid ^2}{\mid M_{RR} \mid ^2} 
\]
and 
\[
R_{R\rightarrow L} \hspace{.1cm} T^*_{R\rightarrow L} + R^*_{R\rightarrow L} 
\hspace{.1cm} T_{R\rightarrow L} = - 2 \hspace{.1cm} Re(M_{RL})\,. 
\]

The two reflection amplitudes are related by the non-diagonal $M$-matrix
elements 
\[
\frac{R_{R\rightarrow L}}{R_{L\rightarrow R}} = - \frac{M_{RL}}{M_{LR}}\,. 
\]
It is also easy to check that 
\[
\det M \equiv M_{RR} M_{LL} - M_{RL} M_{LR} = 1\,, 
\]
which amounts to $T_{L\rightarrow R}=T_{R\rightarrow L}$.

It is worthwhile to spend a few words on the behaviour of the transmission
and reflection coefficients under the exchange of the flux generating part
of the potential, $V(x)=-V_0+iV_1$, with the flux absorbing part, $%
V(x)=-V_0-iV_1$. In our notations, the exchange $V_1\leftrightarrow -V_1$ is
equivalent to $\alpha_0\leftrightarrow \alpha_1$. Since we have already
noticed that $M_{RR}$ is symmetric under that exchange, so is $%
T_{L\rightarrow R}=1/M_{RR}$ and, consequently, $T_{R\rightarrow
L}=T_{L\rightarrow R}$.

The reflection coefficients have a different behaviour: from the relation 
\[
M_{LR}(\alpha_1,\alpha_0,b,k) =-M_{RL}(\alpha_0,\alpha_1,b,k)\,, 
\]
we immediately deduce that 
\[
R_{R\rightarrow L}(\alpha_0,\alpha_1,b,k)= R_{L\rightarrow
R}(\alpha_1,\alpha_0,b,k)\,. 
\]

This behaviour of reflection coefficients of $\mathcal{PT}$-symmetric
potentials was first noticed in Ref.~\cite{Ah04} and discussed there in
detail for the parallel case of the square barrier, as well as of the
hyperbolic Scarf potential, to be treated in subsection \ref{Scarf}.

\subsection{Multiple Square Well}

In order to extend the formalism of the preceding section to a multiple
potential well, it is convenient to consider the change of the $M$ matrix
elements in the case the single well is not centred on the origin, but on an
arbitrary point $X_0$ on the real axis. Under the coordinate shift $%
x^{\prime}=x-X_0$, the basic vectors are changed as follows 
\[
\left|R^{\prime}\right\rangle=e^{-ikX_0}\left|R\right\rangle; \qquad
\left|L^{\prime}\right\rangle=e^{ikX_0}\left|L\right\rangle\,, 
\]
and formula (\ref{C_shift}) holds, with a shift $C=-X_0$, so that the
diagonal elements of the $M$ matrix remain unchanged, while the off-diagonal
ones are changed by a phase factor 
\[
\left( 
\begin{array}{lr}
M^{\prime}_{RR} & M^{\prime}_{RL} \\ 
M^{\prime}_{LR} & M^{\prime}_{LL}
\end{array}
\right) = \left( 
\begin{array}{lr}
M_{RR} & e^{-2ikX_0}M_{RL} \\ 
e^{2ikX_0}M_{LR} & M_{LL}
\end{array}
\right). 
\]

We are thus ready to solve the simplest problem of multiple wells, i. e. two
identical square wells of width $2b$ separated by a zero-potential region of
width $2a$ centered on the origin. The $M$ matrix of this problem is nothing
but the product of the $M$ matrices of the two wells, with off-diagonal
elements properly changed in phase so as to take into account the shifts of
the well centres with respect to the origin 
\begin{equation}
M^{(dw)} = M^{(1)}\cdot M^{(2)}= \left( 
\begin{array}{lr}
M_{RR} & e^{2ik(a+b)}M_{RL} \\ 
e^{-2ik(a+b)}M_{LR} & M_{LL}
\end{array}
\right) \left( 
\begin{array}{lr}
M_{RR} & e^{-2ik(a+b)}M_{RL} \\ 
e^{+2ik(a+b)}M_{LR} & M_{LL}
\end{array}
\right)\,.  \label{M_12}
\end{equation}
The $M_{ij}$ elements are, of course, the same as in the preceding section.

In order to extend the above formalism to an arbitrary number, $n$, of
identical square wells of width $2b$ separated by intervals of constant
length $2a$, it is convenient to introduce a new transfer matrix, $T$,
connected with $M$ as follows 
\begin{equation}
\left( 
\begin{array}{lr}
T_{RR} & T_{RL} \\ 
T_{LR} & T_{LL}
\end{array}
\right) = \left( 
\begin{array}{lr}
M_{RR}e^{-2ik(a+b)} & M_{RL}e^{2ika} \\ 
M_{LR}e^{-2ika} & M_{LL}e^{2ik(a+b)}
\end{array}
\right)\,.  \label{T_mat}
\end{equation}

It is now easy to check that the $M^{(i)}$ matrices of formula (\ref{M_12})
can be written in terms of $T$ in the following way 
\[
M^{(1)}=\mathcal{D}^*(-a-2b)\cdot T\cdot\mathcal{D}(a),\qquad M^{(2)}=%
\mathcal{D}^*(a)\cdot T\cdot\mathcal{D}(a+2(a+b))\,, 
\]
where $\mathcal{D}(x)$ is the diagonal matrix defined by formula (\ref{Shift}%
). It is worth pointing out that the argument $u_1\equiv -a-2b$ of the first 
$\mathcal{D}^*$ matrix is the coordinate of the left edge of the first well,
while the argument $v_1 \equiv u_1+2(a+b)=a$ of the first $\mathcal{D}$
matrix corresponds to the left edge of the second well, obtained by summing
to the left edge of the first well the spatial period $2(a+b)$. A similar
interpretation holds for the arguments $u_2\equiv a$ and $v_2\equiv a+2(a+b)$
of the second well.

Therefore, the $M$ matrix of the double square well can be written in the
very simple form 
\[
M=\mathcal{D}^*(-a-2b)\cdot T\cdot\mathcal{D}(a)\cdot\mathcal{D}^*(a)\cdot
T\cdot\mathcal{D}(a+2(a+b)) =\mathcal{D}^*(-a-2b)\cdot T^2\cdot \mathcal{D}%
(a+2(a+b))\,. 
\]
The generalization to an arbitrary number, $n$ , of square wells with space
period $s\equiv 2(a+b)$ over an interval of length $L=ns$ is trivial 
\[
M=\mathcal{D}^*(u_1)\cdot T^n \cdot \mathcal{D}(u_1+ns)\,. 
\]

In the $n\rightarrow\infty$ limit, the above formula corresponds to the $%
\mathcal{PT}$-symmetric version of a model of one-dimensional crystal, and
might be used in an analysis similar to that already carried out in Ref.~%
\cite{BH02} in the hermitian case.

\subsection{The Hyperbolic Scarf Potential}

\label{Scarf}

As an another example of solvable potential, we consider the hyperbolic
Scarf (or Scarf II) potential, whose scattering solutions were investigated
in Ref. \cite{KS88} in the hermitian case and in Ref.\cite{Le01} in the $%
\mathcal{PT}$-symmetric case. This potential allows simple analytic
solutions for the transmission and reflection coefficients, thus displaying
explicitly the singularity structure in the complex $k$ plane characteristic
of a $\mathcal{PT}$-symmetric potential. General comments on these
singularities as poles of the $S$ matrix connected with bound states and
resonances were recently made in ref.~\cite{Mu04}. We repeat here the
hermitian case of the Scarf II potential, because the solutions given in
Ref. \cite{KS88} contain several misprints. Following the notations of Ref.~%
\cite{Le01}, the hermitian Scarf II potential is written in the form 
\begin{equation}
V\left( x\right) =\left( \lambda ^2-s\left( s+1\right) \right) \frac 1{\cosh
^2x}+\lambda \left( 2s+1\right) \frac{\sinh x}{\cosh ^2x}\,,  \label{ScII}
\end{equation}
where $\lambda $ and $s$ are real parameters. Two independent scattering
solutions are: 
\begin{equation}
F_1\left( x\right) =\left( 1+iy\right) ^{-\frac{s-i\lambda }2}\left(
1-iy\right) ^{-\frac{s+i\lambda }2}F\left( -s-ik,-s+ik,i\lambda -s+\frac 12;%
\frac{1+iy}2\right)\, ;  \label{F_1}
\end{equation}
\begin{equation}
F_2\left( x\right) =\left( 1+iy\right) ^{\frac{s+1-i\lambda }2}\left(
1-iy\right) ^{-\frac{s+i\lambda }2}F\left( \frac 12-i\lambda -ik,\frac
12-i\lambda +ik,s+\frac 32-i\lambda ;\frac{1+iy}2\right)\, ,  \label{F_2}
\end{equation}
where $y\equiv \sinh x$ and $F\left( a,b,c;t\right) $ is the hypergeometric
function. By exploiting the following asymptotic formula of the
hypergeometric function~\cite{As72} 
\[
\lim_{\left| t\right| \rightarrow \infty }F\left( a,b,c;t\right) =\Gamma
\left( c\right) \left( \frac{\Gamma \left( b-a\right) }{\Gamma \left(
b\right) \Gamma \left( c-a\right) }\left( -t\right) ^{-a}+\frac{\Gamma
\left( a-b\right) }{\Gamma \left( a\right) \Gamma \left( c-b\right) }\left(
-t\right) ^{-b}\right) 
\]
and the elementary limit $\lim_{x\rightarrow \pm \infty }\sinh x=\pm \frac{%
\exp \left( \pm x\right) }2$, we readily obtain the behaviour of $F_1$ and $%
F_2$ at $x\rightarrow \pm \infty $ : 
\[
\lim_{x\rightarrow +\infty }F_1\left( x\right)
=a_{1+}e^{ikx}+b_{1+}e^{-ikx}\,, 
\]
with 
\begin{equation}
a_{1+}=\frac{e^{-\frac \pi 2\left( \lambda -k+is\right) }}{2^{s+2ik}}\frac{%
\Gamma \left( i\lambda -s+\frac 12\right) \Gamma \left( 2ik\right) }{\Gamma
\left( -s+ik\right) \Gamma \left( \frac 12+i\lambda +ik\right) }\,;
\label{a1+}
\end{equation}
\begin{equation}
b_{1+}=\frac{e^{-\frac \pi 2\left( \lambda +k+is\right) }}{2^{s-2ik}}\frac{%
\Gamma \left( i\lambda -s+\frac 12\right) \Gamma \left( -2ik\right) }{\Gamma
\left( -s-ik\right) \Gamma \left( \frac 12+i\lambda -ik\right) }.
\label{b1+}
\end{equation}
\[
\lim_{x\rightarrow -\infty }F_1\left( x\right)
=a_{1-}e^{ikx}+b_{1-}e^{-ikx}\,, 
\]
with 
\begin{equation}
a_{1-}=\frac{e^{\frac \pi 2\left( \lambda +k+is\right) }}{2^{s-2ik}}\frac{%
\Gamma \left( i\lambda -s+\frac 12\right) \Gamma \left( -2ik\right) }{\Gamma
\left( -s-ik\right) \Gamma \left( \frac 12+i\lambda -ik\right) };
\label{a1-}
\end{equation}
\begin{equation}
b_{1-}=\frac{e^{\frac \pi 2\left( \lambda -k+is\right) }}{2^{s+2ik}}\frac{%
\Gamma \left( i\lambda -s+\frac 12\right) \Gamma \left( 2ik\right) }{\Gamma
\left( -s+ik\right) \Gamma \left( \frac 12+i\lambda +ik\right) }.
\label{b1-}
\end{equation}

Therefore, the following relations hold 
\begin{eqnarray*}
a_{1-} &=&e^{\pi \left( \lambda +k+is\right) }b_{1+}; \\
b_{1-} &=&e^{\pi \left( \lambda -k+is\right) }a_{1+}.
\end{eqnarray*}

The procedure is repeated for the second solution, $F_2$, with the following
results: 
\[
\lim_{x\rightarrow +\infty }F_2\left( x\right)
=a_{2+}e^{ikx}+b_{2+}e^{-ikx}\,, 
\]
where 
\begin{equation}
a_{2+}=\frac{e^{\frac \pi 2\left( \lambda +k+i\left( s+1\right) \right) }}{%
2^{2ik+i\lambda -\frac 12}}\frac{\Gamma \left( s+\frac 32-i\lambda \right)
\Gamma \left( 2ik\right) }{\Gamma \left( \frac 12-i\lambda +ik\right) \Gamma
\left( s+1+ik\right) }\,;  \label{a2+}
\end{equation}
\begin{equation}
b_{2+}=\frac{e^{\frac \pi 2\left( \lambda -k+i\left( s+1\right) \right) }}{%
2^{-2ik+i\lambda -\frac 12}}\frac{\Gamma \left( s+\frac 32-i\lambda \right)
\Gamma \left( -2ik\right) }{\Gamma \left( \frac 12-i\lambda -ik\right)
\Gamma \left( s+1-ik\right) }\,.  \label{b2+}
\end{equation}
\[
\lim_{x\rightarrow -\infty }F_2\left( x\right)
=a_{2-}e^{ikx}+b_{2-}e^{-ikx}\,, 
\]
where 
\begin{equation}
a_{2-}=\frac{e^{-\frac \pi 2\left( \lambda -k+i\left( s+1\right) \right) }}{%
2^{-2ik+i\lambda -\frac 12}}\frac{\Gamma \left( s+\frac 32-i\lambda \right)
\Gamma \left( -2ik\right) }{\Gamma \left( \frac 12-i\lambda -ik\right)
\Gamma \left( s+1-ik\right) }\,;  \label{a2-}
\end{equation}
\begin{equation}
b_{2-}=\frac{e^{-\frac \pi 2\left( \lambda +k+i\left( s+1\right) \right) }}{%
2^{2ik+i\lambda -\frac 12}}\frac{\Gamma \left( s+\frac 32-i\lambda \right)
\Gamma \left( 2ik\right) }{\Gamma \left( \frac 12-i\lambda +ik\right) \Gamma
\left( s+1+ik\right) }\,.  \label{b2-}
\end{equation}

Therefore: 
\begin{eqnarray*}
a_{2-} &=&e^{-\pi \left( \lambda -k+i\left( s+1\right) \right) }b_{2+}; \\
b_{2-} &=&e^{-\pi \left( \lambda +k+i\left( s+1\right) \right) }a_{2+}.
\end{eqnarray*}

It is worthwhile to stress that the relations connecting $a_{i-}$ with $%
b_{i+}$ and $b_{i-}$ with $a_{i+}$ ($i=1,2$) are valid not only for the
hermitian potential (real $s$ and $\lambda $), but also for the $\mathcal{PT}
$-symmetric potential with real $s$ and imaginary $\lambda =i\lambda
^{\prime }$, provided the $x$ coordinate is real.

After some manipulations of the $\Gamma $ functions in the asymptotic
amplitudes, it is not difficult to obtain the compact form of the $T_{L\rightarrow R}$ and $%
R_{L\rightarrow R}$ coefficients first derived in Ref.\cite{KS88}, corrected in Ref.\cite
{Le01} for a wrong sign, and repeated here for the sake of completeness: 
\begin{eqnarray}
T_{L\rightarrow R} &=&\frac{\Gamma \left( -s-ik\right) \Gamma \left(
s+1-ik\right) \Gamma \left( \frac 12+i\lambda -ik\right) \Gamma \left( \frac
12-i\lambda -ik\right) }{\Gamma \left( -ik\right) \Gamma \left( 1-ik\right)
\left( \Gamma \left( \frac 12-ik\right) \right) ^2}\,;  \label{TR_L} \\
R_{L\rightarrow R} &=&T_{L\rightarrow R}\left( \frac{\cos \left( \pi
s\right) \sinh \left( \pi \lambda \right) }{\cosh \left( \pi k\right) }+i%
\frac{\sin \left( \pi s\right) \cosh \left( \pi \lambda \right) }{\sinh
\left( \pi k\right) }\right) \,.
\end{eqnarray}

Formulae (\ref{TR_L}) hold also for the $\mathcal{PT}$-symmetric version of
the potential with real $s$ and imaginary $\lambda =i\lambda ^{\prime }$. In
this case the hyperbolic functions of $\lambda $ are changed into circular
functions of $\lambda ^{\prime }$: $\sinh (i\lambda ^{\prime })=i\sin
(\lambda ^{\prime })$, $\cosh (i\lambda ^{\prime })=\cos (\lambda ^{\prime })
$. Formulae (\ref{TR_L}) satisfy the unitarity condition 
\begin{equation}
\left| T_{L\rightarrow R}\right| ^2+\left| R_{L\rightarrow R}\right| ^2=1\,.
\label{Uc}
\end{equation}
in the hermitian case (real $\lambda $), while in the $\mathcal{PT}$%
-symmetric case (imaginary $\lambda $) unitarity may be broken : for
instance, when $s$ is integer and $\lambda /i$ half-integer, $\left|
T_{L\rightarrow R}\right| ^2+\left| R_{L\rightarrow R}\right| ^2\rightarrow
\infty $ when $k\rightarrow 0$ .

By using the general definitions of transmission and reflection
coefficients, considered as functions of the coupling strength $\lambda $,
together with the relations connecting $a_{i\pm }$ and $b_{i\pm }$ ($i=1,2$%
), it is easy to check that $T_{R\rightarrow L}(\lambda )=T_{L\rightarrow
R}(\lambda )$ and $R_{R\rightarrow L}(\lambda )=R_{L\rightarrow R}(-\lambda
) $ in both the hermitian and the $\mathcal{PT}$-symmetric case mentioned
above, as already noticed in Ref.\cite{Ah04}.

A further transformation preserving $\mathcal{PT}$ symmetry of the Scarf
potential with real $s$ and imaginary $\lambda $ is the complex coordinate
shift $x\rightarrow x+i\epsilon $ ($-\frac \pi 2<\epsilon <+\frac \pi 2$, in
order to avoid singularities in the potential).

The asymptotic forms of the two independent solutions $F_1$ and $F_2$ are
easily computed by the procedure described in the previous pages: 
\begin{eqnarray*}
\lim_{x\rightarrow +\infty }F_1\left( x+i\epsilon \right)
&=&\lim_{x\rightarrow +\infty }\left( i\frac{e^{x+i\epsilon }}2\right) ^{-%
\frac{s-i\lambda }2}\left( -i\frac{e^{x+i\epsilon }}2\right) ^{-\frac{%
s+i\lambda }2} \\
&&\cdot F\left( -s-ik,-s+ik,i\lambda -s+\frac 12;i\frac{e^{x+i\epsilon }}%
4\right) \\
&=&a_{1+}\left( \epsilon \right) e^{ikx}+b_{1+}\left( \epsilon \right)
e^{-ikx}\,,
\end{eqnarray*}
where 
\begin{equation}
a_{1+}\left( \epsilon \right) =a_{1+}e^{-k\epsilon }\,;  \label{a1+eps}
\end{equation}
\begin{equation}
b_{1+}\left( \epsilon \right) =b_{1+}e^{k\epsilon }\,.  \label{b1+eps}
\end{equation}

Here, $a_{1+}$ and $b_{1+}$ on the right-hand-side of the previous equations
are given by formulae (\ref{a1+}) and (\ref{b1+}), respectively. In the same
way, one gets 
\[
\lim_{x\rightarrow -\infty }F_1\left( x+i\epsilon \right) =a_{1-}\left(
\epsilon \right) e^{ikx}+b_{1-}\left( \epsilon \right) e^{-ikx}\,, 
\]
with 
\begin{equation}
a_{1-}\left( \epsilon \right) =a_{1-}e^{-k\epsilon }\,;  \label{a1-eps}
\end{equation}
\begin{equation}
b_{1-}\left( \epsilon \right) =b_{1-}e^{k\epsilon }.  \label{b1-eps}
\end{equation}

Here, $a_{1-}$ and $b_{1-}$ on the r.h.s. are obviously given by formulae (%
\ref{a1-}) and (\ref{b1-}), respectively.

The two limits of $F_2\left( x+i\epsilon \right) $ are computed in the same
way 
\[
\lim_{x\rightarrow \pm \infty }F_2\left( x+i\epsilon \right) =a_{2\pm
}\left( \epsilon \right) e^{ikx}+b_{2\pm }\left( \epsilon \right)
e^{-ikx}\,, 
\]
with similar results: 
\begin{equation}
a_{2+}\left( \epsilon \right) =a_{2+}e^{-k\epsilon }\,;  \label{a2+eps}
\end{equation}
\begin{equation}
b_{2+}\left( \epsilon \right) =b_{2+}e^{k\epsilon }\,;  \label{b2+eps}
\end{equation}
\begin{equation}
a_{2-}\left( \epsilon \right) =a_{2-}e^{-k\epsilon }\,;  \label{a2-eps}
\end{equation}
\begin{equation}
b_{2-}\left( \epsilon \right) =b_{2-}e^{k\epsilon }\,.  \label{b2-eps}
\end{equation}

It is thus immediate to check the result of Ref.\cite{Le01}: 
\begin{equation}
T_{L\rightarrow R}\left( \epsilon,\lambda \right) = T_{L\rightarrow
R}(0,\lambda)\,;  \label{T_Leps}
\end{equation}
\begin{equation}
R_{L\rightarrow R}\left( \epsilon, \lambda \right) = R_{L\rightarrow
R}(0,\lambda)e^{2k\epsilon }\,,  \label{R_Leps}
\end{equation}
where the $T_{L\rightarrow R}$ and $R_{L\rightarrow R}$ coefficients on the
r.h.s. of formulae (\ref{T_Leps}-\ref{R_Leps}) are given by (\ref{TR_L}).
Unitarity is obviously broken by $\epsilon \neq 0$. Moreover, it easy to
check that, in the same case 
\begin{equation}
T_{R\rightarrow L}\left( \epsilon, \lambda \right) = T_{L\rightarrow
R}\left( \epsilon, \lambda \right)\, ;  \label{T_Reps}
\end{equation}
\begin{equation}
R_{R\rightarrow L}\left( \epsilon,\lambda \right) = R_{L\rightarrow R}\left(
\epsilon,-\lambda \right) e^{-4k\epsilon }\,.  \label{R_Reps}
\end{equation}

\subsection{Reflectionless Potentials}

\label{refless} Hermitian reflectionless potentials are much studied in the
literature\cite{KM56, KR86, Ma05}. In this section, we discuss two examples
of reflectionless $\mathcal{PT}$-symmetric potentials. The first example is
the regularized one-dimensional form of the ''centrifugal'' potential 
\begin{equation}
V\left( x\right) =\frac \alpha {\left( x+i\varepsilon \right) ^2}\,,
\label{centrifugal}
\end{equation}
where $\alpha $ is a real strength and $\varepsilon $ a real constant that
removes the singularity at the origin. The time-independent Schr\H{o}dinger
equation for the potential under investigation reads, in units $\hbar =2m=1$%
\begin{equation}
\left( -\frac{d^2}{dx^2}+\frac \alpha {\left( x+i\varepsilon \right)
^2}\right) \Psi =k^2\Psi \;,  \label{Schr_eq_c}
\end{equation}

We introduce the complex variable $z=k\left( x+i\varepsilon \right) $ and
express Eq. (\ref{Schr_eq_c}) in terms of $z$, 
\begin{equation}
z^2\frac{d^2}{dz^2}\Psi +\left( z^2-\alpha \right) \Psi =0\;.
\label{Schr_eq_c2}
\end{equation}

If we define the new function $\Phi \left( z\right) $ such that $\Psi \left(
z\right) =z^{1/2}\Phi \left( z\right) $, the equation fulfilled by $\Phi $
is promptly obtained from Eq.(\ref{Schr_eq_c2}) in the form 
\begin{equation}
z^2\frac{d^2}{dz^2}\Phi +z\frac d{dz}\Phi +\left( z^2-\alpha -\frac
14\right) \Phi =0\;,  \label{Bessel}
\end{equation}

$i.e.$ a Bessel equation with square index $\nu ^2=\alpha +1/4$.

A couple of linearly independent solutions to Eq.(\ref{Bessel}) with the
appropriate asymptotic behaviour for $\Psi $ to be a scattering solution to
Eq.(\ref{Schr_eq_c2}) is provided by the Hankel functions of first and
second type, $H_\nu ^{\left( 1\right) }\left( z\right) $ and $H_\nu ^{\left(
2\right) }\left( z\right) $, respectively, whose lowest order asymptotic
expansions are \cite{RG65} 
\begin{eqnarray}
\lim_{\left| z\right| \rightarrow \infty }H_\nu ^{\left( 1\right) }\left(
z\right) &=&\left( \frac 2{\pi z}\right) ^{1/2}\exp \left[ i\left( z-\frac
\pi 2\nu -\frac \pi 4\right) \right] \;,  \label{Hankel} \\
\lim_{\left| z\right| \rightarrow \infty }H_\nu ^{\left( 2\right) }\left(
z\right) &=&\left( \frac 2{\pi z}\right) ^{1/2}\exp \left[ -i\left( z-\frac
\pi 2\nu -\frac \pi 4\right) \right] \;,
\end{eqnarray}

valid for $\Re \nu >-1/2$, $\left| \arg z\right| <\pi $.

The corresponding asymptotic solutions to Eq.(\ref{Schr_eq_c2}) thus are 
\begin{eqnarray}
\lim_{x\rightarrow \infty }\Psi _1\left( x\right) &=&\exp \left(
ikx-k\varepsilon -i\frac \pi 2\nu -i\frac \pi 4\right) \;,  \label{Psi_1_2}
\\
\lim_{x\rightarrow \infty }\Psi _2\left( x\right) &=&\exp \left(
-ikx+k\varepsilon +i\frac \pi 2\nu +i\frac \pi 4\right)
\end{eqnarray}

If the above asymptotic wave functions are written in the general form 
\begin{equation}
\lim_{x\rightarrow \pm \infty }\Psi _i\left( x\right) =a_{i\pm }\exp \left(
ikx\right) +b_{i\pm }\exp \left( -ikx\right) \;,
\end{equation}

we immediately obtain 
\begin{eqnarray}
a_{1+} &=&a_{1-}=\exp \left( -k\varepsilon -i\frac \pi 2\nu -i\frac \pi
4\right) \;;\quad b_{1+}=b_{1-}=0\;.  \label{Cf_ampl} \\
a_{2+} &=&a_{2-}=0\;;\quad b_{2+}=b_{2-}=\exp \left( +k\varepsilon +i\frac
\pi 2\nu +i\frac \pi 4\right) \;.
\end{eqnarray}

Formulae (\ref{Cf_ampl}) show that $\mathcal{PT}$ is an exact symmetry for
this potential. In fact, remembering Eqs. (\ref{PT_ex_0}-\ref{PT_ex_2}), we
have in this case $A_{+}=a_{1+}$, $A_{-}=a_{1-}=A_{+}$, so that $%
A_{+}^{*}/A_{-}=a_{1+}^{*}/a_{1+}=\exp \left( i\pi \left( \nu +1/2\right)
\right) $. In the same way, we get $\widetilde{B}_{+}=b_{2+}$, $\widetilde{B}%
_{-}=b_{2-}=\widetilde{B}_{+}$, and $\widetilde{B}_{+}^{*}/\widetilde{B}%
_{-}=b_{2+}^{*}/b_{2+}=\exp \left( -i\pi \left( \nu +1/2\right) \right) $.

The resulting transmission and reflection coefficients for waves travelling
from left to right and vice versa are promptly evaluated from their
definition 
\begin{eqnarray}
T_{L\rightarrow R} &=&\frac{a_{2+}b_{1+}-a_{1+}b_{2+}}{%
a_{2-}b_{1+}-a_{1-}b_{2+}}=1\;,  \label{T_LR} \\
R_{L\rightarrow R} &=&\frac{b_{1+}b_{2-}-b_{1-}b_{2+}}{%
a_{2-}b_{1+}-a_{1-}b_{2+}}=0\;, \\
T_{R\rightarrow L} &=&\frac{a_{2-}b_{1-}-a_{1-}b_{2-}}{%
a_{2-}b_{1+}-a_{1-}b_{2+}}=1\;, \\
R_{R\rightarrow L} &=&\frac{a_{1+}a_{2-}-a_{1-}a_{2+}}{%
a_{2-}b_{1+}-a_{1-}b_{2+}}=0\;.
\end{eqnarray}

The second example is the hyperbolic Scarf potential with integer coupling
strengths, $s=n$ and $\lambda =im$ . The formulae of the preceding section
read in this case 
\begin{eqnarray}
T_{L\rightarrow R} &=&T_{R\rightarrow L}=(-1)^{n+m}\frac{(n-ik)...(1-ik)}{%
(n+ik)...(1+ik)}\frac{\left( m-\frac 12-ik\right) ...\left( \frac
12-ik\right) }{\left( m-\frac 12+ik\right) ...\left( \frac 12+ik\right) }%
\;,\;  \label{T_RL_PT} \\
R_{L\rightarrow R} &=&R_{L\rightarrow R}=0\;.  \label{R_RL_PT}
\end{eqnarray}
with $\left| T_{_{L\rightarrow R}}\right| =1$. Equations (\ref{T_RL_PT}-\ref
{R_RL_PT}) are sufficient to ensure that the asymptotic wave functions are
eigenstates of $\mathcal{PT}$.

Moreover, the potential possesses bound states, which, in the present
parametrization, turn out to be eigenfunctions of $\mathcal{PT}$ , as
discussed in Refs.\cite{Ah01, Le02b}.

In the Hermitian case, the hyperbolic Scarf potential is reflectionless only
when $s=n$ and $\lambda =0$, corresponding to the well-known case of the P\H{%
o}schl-Teller potential with integer coupling strength.

\subsection{A Non-Local Potential}

\label{nonloc}

Let us go back to the general Schr\"{o}dinger equation (\ref{Schreqt}) for a
monochromatic wave (\ref{psi_t}) of energy $E=k^2$ 
\begin{equation}
-\frac{d^2}{dx^2}\Psi (x)+\lambda \int K(x,y)\Psi (y)dy=k^2\Psi (x)\,,
\label{Schrnl}
\end{equation}
where the potential strength, $\lambda $, is a real number. It is easy to
check, by calculating scalar products, that the kernel of a hermitian
non-local potential satisfies the condition 
\begin{equation}
K(x,y)=K^{*}(y,x)\,.  \label{hermnl}
\end{equation}

In the $L-R$ basis, $K$ is written as a $2\times 2$ hermitian matrix, as a
consequence of the above constraint (\ref{hermnl}) 
\[
K\equiv \left( 
\begin{array}{lr}
K_{RR} & K_{RL} \\ 
K_{LR} & K_{LL}
\end{array}
\right) =\left( 
\begin{array}{lr}
K_{RR}^{*} & K_{LR}^{*} \\ 
K_{RL}^{*} & K_{LL}^{*}
\end{array}
\right) \,\equiv K^{\dagger }. 
\]

Parity invariance of the potential could be similarly checked to imply 
\begin{equation}
K(x,y)=K(-x,-y)\;.  \label{Parinvnl}
\end{equation}
In the $L-R$ basis, this corresponds to 
\begin{equation}
K\equiv \left( 
\begin{array}{lr}
K_{RR} & K_{RL} \\ 
K_{LR} & K_{LL}
\end{array}
\right) =\left( 
\begin{array}{ll}
K_{LL} & K_{LR} \\ 
K_{RL} & K_{RR}
\end{array}
\right) \equiv \mathcal{P}K\mathcal{P}^{-1}\equiv K_{\mathcal{P}}\;,
\end{equation}
as in Section \ref{par}.

The condition of time reversal invariance of $K$ is obtained by using the
definitions of Section \ref{T_inv} in the form 
\begin{equation}
K_{\mathcal{T}}\equiv \mathcal{T}K\mathcal{T}^{-1}=\mathcal{P}K^{*}\mathcal{%
P=}K\;,  \label{Tinvar}
\end{equation}
or, in matrix notation, 
\begin{equation}
\left( 
\begin{array}{lr}
K_{RR} & K_{RL} \\ 
K_{LR} & K_{LL}
\end{array}
\right) =\left( 
\begin{array}{ll}
K_{LL}^{*} & K_{LR}^{*} \\ 
K_{RL}^{*} & K_{RR}^{*}
\end{array}
\right) \;,
\end{equation}
which corresponds to 
\begin{equation}
K(x,y)=K^{*}(x,y)\;.  \label{Tinvnl}
\end{equation}

Conditions similar to those of Section \ref{PTsymm}, therefore, lead us to
introduce $\mathcal{PT}$ invariance of $K$ in the form 
\begin{equation}
K_{\mathcal{PT}}\equiv \mathcal{PT}K\mathcal{T}^{-1}\mathcal{P}^{-1}=\left( 
\begin{array}{lr}
K_{RR}^{*} & K_{RL}^{*} \\ 
K_{LR}^{*} & K_{LL}^{*}
\end{array}
\right) =\left( 
\begin{array}{lr}
K_{RR} & K_{RL} \\ 
K_{LR} & K_{LL}
\end{array}
\right) =K\;.  \label{PTinvar}
\end{equation}

This corresponds to 
\begin{equation}
K(x,y)=K^{*}(-x,-y)\;,  \label{ptnl}
\end{equation}
in agreement with formula (3) of Ref.\cite{Ru05} , which corrects a misprint
in the corresponding formula (113) of Ref.\cite{Mu04}.

In order to deal with a solvable $\mathcal{PT}$-symmetric potential, we
consider only separable kernels of the kind 
\begin{equation}
K(x,y)=g(x)e^{i\alpha x}h(y)e^{i\beta y}\,,  \label{K_pm}
\end{equation}
where $\alpha $ and $\beta $ are real numbers, and $g(x)$ and $h(y)$ are
real functions of their argument, suitably vanishing at $\pm \infty $.

For this kind of kernel, the hermiticity condition (\ref{hermnl}) implies $%
\alpha =-\beta $ and $g=h$. Parity invariance (\ref{Parinvnl}) requires $%
\alpha =\beta =0$ and $g\left( x\right) =g\left( -x\right) $, $h\left(
x\right) =h\left( -x\right) $. Time reversal invariance (\ref{Tinvnl})
requires $\alpha =\beta =0$ , but does not impose conditions on $g$ and $h$.

The various conditions that can be imposed on kernel (\ref{K_pm}) are
summarized in Table I. 
\begin{table}[tbp]
\begin{center}
\begin{tabular}{|c|c|}
\hline
Reality & $\alpha =\beta =0$ \\ \hline
Symmetry under $x\leftrightarrow y$ & $\alpha =\beta $, $g=h$ \\ \hline
Hermiticity & $\alpha =-\beta $, $g=h$ \\ \hline
$\mathcal{P}$ Invariance & $\alpha =\beta =0$, $g\left( x\right) =g\left(
-x\right) $, $h\left( y\right) =h\left( -y\right) $ \\ \hline
$\mathcal{T}$ Invariance & $\alpha =\beta =0$ \\ \hline
$\mathcal{PT}$ Invariance & $g\left( x\right) =g\left( -x\right) $, $h\left(
y\right) =h\left( -y\right) $ \\ \hline
\end{tabular}
\end{center}
\caption{Possible symmetries of the separable kernel (\ref{K_pm})}
\end{table}
Finally, $\mathcal{PT}$ invariance (\ref{ptnl}) does not impose conditions
on $\alpha $ and $\beta $, but requires $g(x)=g(-x)\,,h(y)=h(-y)\,$. As an
important consequence, their Fourier transforms, $\tilde{g}(q)$ and $\tilde{h%
}(q^{\prime })$, are real even functions, too. In order to solve eq. (\ref
{Schrnl}), we resort to the Green's function method.
 As is known, the Green's function of the problem is a
solution to Eq. (\ref{Schrnl}) with the potential term replaced with a Dirac
delta function 
\begin{equation}
\frac{d^2}{dx^2}G_{\pm }(x,y)+(k^2\pm i\varepsilon )G_{\pm }(x,y)=\delta
(x-y)\,.  \label{Helmeq}
\end{equation}

Here, we introduce the infinitesimal positive number $\varepsilon$ in order
to shift upwards, or downwards in the complex momentum plane the
singularities of the Fourier transform of the Green's function, $%
G_{\pm}(q,q^{\prime})$, lying on the real axis.

In fact, after defining the Fourier transform, $\tilde{f}(q)$, of a generic
function $f(x)$ as 
\[
\tilde{f}(q)=\int_{-\infty}^{+\infty}f(x)e^{-iqx}dx\quad \leftrightarrow
\quad f(x)=\frac{1}{2\pi} \int_{-\infty}^{+\infty}\tilde{f}(q)e^{iqx}dq \,, 
\]
and expressing $G_{\pm}(x,y)$ and $\delta(x-y)$ in terms of their Fourier
transforms, we quickly solve eq. (\ref{Helmeq}) for $G_{\pm}$ 
\[
\tilde{G}_\pm(q,q^{\prime})=\frac{2\pi\delta(q+q^{\prime})}{-q^2+k^2\pm
i\varepsilon}, 
\]

Therefore, the Green's function in coordinate space is 
\begin{equation}
G_{\pm}(x,y)=-\frac{1}{2\pi}\int_{-\infty}^{+\infty}\frac{1}{q^2-k^2\mp
i\varepsilon}e^{iq(x-y)}dq=G_{\pm}(x-y)\,.  \label{Greenf}
\end{equation}
The integral (\ref{Greenf}) is easily computed by the method of residues.
 In fact, the integrand in $G_+(x-y)$
has two first order poles at $q_1=k+i\varepsilon^{\prime}$ and $%
q_2=-k-i\varepsilon^{\prime}$, where $\varepsilon^{\prime}=\varepsilon/(2k)$%
: the integral is thus computed along a contour made of the real $q$ axis
and of a half-circle of infinite radius in the upper half-plane for $x-y>0$,
on which the integrand vanishes, thus enclosing the pole at $q=q_1$, and in
the lower half-plane for $x-y<0$, enclosing the pole at $q=q_2$, for the
same reason. Notice that the $G_+$ contour integral is done in the
counterclockwise direction for $x-y>0$, while it is done in the clockwise
direction for $x-y<0$, so that the latter acquires a global sign opposite to
the former. The result is 
\begin{equation}
G_+(x-y)=-\frac{i}{2k}\left[e^{ik(x-y)}\theta(x-y)+e^{-ik(x-y)}\theta(y-x)%
\right]\,,  \label{G_p}
\end{equation}
where $\theta(x)$ is the step function, equal to 1 for $x>0$ and 0 otherwise.

The second Green's function, $G_-(x-y)$, is the complex conjugate of $%
G_+(x-y)$ 
\begin{equation}
G_-(x-y)=\frac{i}{2k}\left[e^{-ik(x-y)}\theta(x-y)+e^{ik(x-y)}\theta(y-x)%
\right]\,.  \label{G_m}
\end{equation}

Now, we go back to eq. (\ref{Schrnl}) with kernel (\ref{K_pm}), call $%
\Psi_\pm(x)$ two linearly independent solutions, for a reason that will
become clear in the next few lines, and define the following integral
depending on $\Psi_\pm$ 
\[
I_\pm(\beta,k)=\int_{-\infty}^{+\infty}e^{i\beta y} h(y)\Psi_\pm(y)dy\,. 
\]
It is easy to show that $I_\pm(\beta,k)$ can be written as a convolution of
the Fourier transforms of $h(y)$ and $\Psi_{\pm}(y)$.

The general solution to eq. (\ref{Schrnl}) is thus implicitly written as 
\begin{equation}
\Psi_\pm (x)= c_\pm e^{ikx}+d_\pm e^{-ikx}+\lambda I_\pm (\beta,k)
\int_{-\infty}^{+\infty}G_\pm (x-y)g(y)e^{i\alpha y}dy\,.  \label{Psipm}
\end{equation}

Eq. (\ref{Psipm}) allows us to express $I_{\pm }(\beta ,k)$ in terms of the
constants $c_{\pm }$ and $d_{\pm }$ and of Fourier transforms of known
functions : in fact, by multiplying both sides by $h(x)e^{i\beta x}$ and
integrating over $x$, we obtain 
\begin{equation}
I_{\pm }(\beta ,k)=c_{\pm }\tilde{h}(k+\beta )+d_{\pm }\tilde{h}(k-\beta
)+\lambda N_{\pm }(\alpha ,\beta ,k)I_{\pm }(\beta ,k)\,,  \label{Ipm1}
\end{equation}
where we have exploited the symmetry $\tilde{h}(-k-\beta )=\tilde{h}(k+\beta
)$ and $N_{\pm }$ is defined as 
\begin{eqnarray}
N_{\pm }(\alpha ,\beta ,k) &=&\int_{-\infty }^{+\infty }h(x)e^{i\beta
x}G_{\pm }(x-y)g(y)e^{i\alpha y}dxdy \\
&=&\mp \frac i{2k}\int_{-\infty }^{+\infty }h(x)e^{i\beta x}e^{\pm ik\left|
x-y\right| }g(y)e^{i\alpha y}dxdy,  \nonumber  \label{Npm}
\end{eqnarray}
so that 
\begin{equation}
I_{\pm }(\beta ,k)=\frac{c_{\pm }\tilde{h}(k+\beta )+d_{\pm }\tilde{h}%
(k-\beta )}{1-\lambda N_{\pm }(\alpha ,\beta ,k)}=(c_{\pm }\tilde{h}(k+\beta
)+d_{\pm }\tilde{h}(k-\beta ))D_{\pm }\,,  \label{Ipm2}
\end{equation}
where 
\[
D_{\pm }(\alpha ,\beta ,k)\equiv \frac 1{1-\lambda N_{\pm }(\alpha ,\beta
,k)}\,.
\]

Let us examine now the asymptotic behaviour of the two independent
solutions, starting from $\Psi_+(x)$ 
\begin{equation}
\Psi_+(x) = c_+ e^{ikx}+d_+ e^{-ikx}+\lambda
I_+(\beta,k)\int_{-\infty}^{+\infty}G_+ (x-y)g(y)e^{i\alpha y}dy \, .
\label{Psi1}
\end{equation}
The asymptotic behaviour of the integral on the r. h. s. of eq. (\ref{Psi1})
is promptly evaluated by observing that, according to eq. (\ref{G_p}), 
\[
\lim_{x\rightarrow \pm \infty}G_+ (x-y)=-\frac{i}{2k}e^{\pm ik(x-y)}\,, 
\]
so that 
\[
\lim_{x\rightarrow \pm \infty}\Psi_+ (x)=c_+ e^{ikx}+d_+e^{-ikx}-i\omega I_+
(\beta,k)\tilde{g}(k\mp \alpha)e^{\pm ikx}\,, 
\]
where we have put $\omega=\lambda/(2k)$.

Remembering the expression (\ref{Ipm2}) of $I_+$, we finally obtain 
\begin{eqnarray*}
\lim_{ x\rightarrow -\infty}\Psi_+ (x) &=& c_+ e^{ikx}+\left\{d_+ -i\omega%
\tilde{g}(k+\alpha) \left[c_+\tilde{h}(k+\beta)+d_+\tilde{h}%
(k-\beta)\right]D_+\right\} e^{-ikx}\,, \\
\lim_{ x\rightarrow +\infty}\Psi_+ (x) &=& \left\{ c_+ -i\omega\tilde{g}%
(k-\alpha) \left[c_+\tilde{h}(k+\beta)+d_+\tilde{h}(k-\beta)\right]D_+\right%
\}e^{ikx} +d_+ e^{-ikx}\,.
\end{eqnarray*}

The constants $c_+$ and $d_+$ are fixed by initial conditions: if we impose
that $\Psi_+(x)$ represents a wave travelling from left to right, according
to formula (\ref{Psi_1}), we immediately have $c_+ =1$, $d_+ =0$ and 
\begin{eqnarray}
T_{L\rightarrow R} &=& 1-i\omega\tilde{g}(k-\alpha)\tilde{h}%
(k+\beta)D_+(\alpha,\beta,k)\,, \\
R_{L\rightarrow R} &=& -i\omega\tilde{g}(k+\alpha)\tilde{h}%
(k+\beta)D_+(\alpha,\beta,k)\,.  \nonumber  \label{TRLR}
\end{eqnarray}

It is worthwhile to point out that the above expressions break unitarity,
i.e. $\mid T_{L\rightarrow R}\mid^2 +\mid R_{L\rightarrow R}\mid^2 \neq 1$,
because probability flux is not conserved in general.

We come now to the second solution, $\Psi_- (x)$, written in the form 
\begin{equation}
\Psi_-(x) = c_- e^{ikx}+d_- e^{-ikx}+\lambda
I_-(\beta,k)\int_{-\infty}^{+\infty}G_- (x-y)g(y)e^{i\alpha y}dy\, .
\label{Psi2}
\end{equation}
The asymptotic behaviour of the Green's function, $G_- (x)$, is now 
\[
\lim_{x\rightarrow \pm \infty}G_- (x-y)=\frac{i}{2k}e^{\mp ik(x-y)}\,, 
\]
so that 
\[
\lim_{x\rightarrow \pm \infty}\Psi_- (x)=c_- e^{ikx}+d_- e^{-ikx}+i\omega
I_- (\beta,k)\tilde{g}(k\pm \alpha)e^{\mp ikx}\,, 
\]
or, using the explicit expression (\ref{Ipm2}) of $I_-$, 
\begin{eqnarray*}
\lim_{ x\rightarrow -\infty}\Psi_- (x) &=& d_- e^{-ikx}+\left\{c_- +i\omega%
\tilde{g}(k-\alpha) \left[c_-\tilde{h}(k+\beta)+d_-\tilde{h}%
(k-\beta)\right]D_-\right\}e^{ikx}\,, \\
\lim_{ x\rightarrow +\infty}\Psi_- (x) &=& c_- e^{ikx}+\left\{d_- +i\omega%
\tilde{g}(k+\alpha) \left[c_-\tilde{h}(k+\beta)+d_-\tilde{h}%
(k-\beta)\right]D_-\right\}e^{-ikx}\,.
\end{eqnarray*}

Since $\Psi_-(x)$ and $\Psi_+(x)$ are linearly independent, we can impose
that $\Psi_-(x)$ is a wave travelling from right to left, according to
formula (\ref{Psi_2}); the initial conditions are 
\begin{eqnarray*}
c_- +i\omega\tilde{g}(k-\alpha)(c_-\tilde{h}(k+\beta)+d_-\tilde{h}%
(k-\beta))D_-(\alpha,\beta,k) &=& 0\,, \\
d_- +i\omega\tilde{g}(k+\alpha)(c_-\tilde{h}(k+\beta)+d_-\tilde{h}%
(k-\beta)))D_-(\alpha,\beta,k)) &=& 1\,,
\end{eqnarray*}
where 
\begin{equation}
d_- = T_{R\rightarrow L},\qquad c_- =R_{R\rightarrow L}\,.  \label{TRRL}
\end{equation}

We then obtain 
\begin{eqnarray}
T_{R\rightarrow L} &=& 1-i\omega\tilde{g}(k+\alpha)\tilde{h}(k-\beta)%
\mathcal{D}_-(\alpha,\beta,k) \,, \\
R_{R\rightarrow L} &=& -i\omega\tilde{g}(k-\alpha)\tilde{h}(k-\beta)\mathcal{%
D}_-(\alpha,\beta,k) \,.  \nonumber  \label{TR_RL}
\end{eqnarray}
where 
\[
\mathcal{D}_-(\alpha,\beta,k) =\frac{1} {1-\lambda N_-+i\omega (\tilde{g}%
(k+\alpha)\tilde{h}(k-\beta) +\tilde{g}(k-\alpha)\tilde{h}(k+\beta))} \,. 
\]

Formulae (\ref{TRLR}-\ref{TRRL}) show that, in general, for a $\mathcal{PT}$%
-symmetric non-local potential, $T_{R\rightarrow L}\neq T_{L\rightarrow R}$.
In fact, from the quoted formulae, 
\[
T_{R\rightarrow L}-T_{L\rightarrow R}=i\omega \Delta D_{+}(\alpha ,\beta ,k)%
\mathcal{D}_{-}(\alpha ,\beta ,k)\,,
\]
where 
\begin{eqnarray*}
\Delta  &=&\tilde{g}(k-\alpha )\tilde{h}(k+\beta )-\tilde{g}(k+\alpha )%
\tilde{h}(k-\beta ) \\
&&+\lambda (N_{+}\tilde{g}(k+\alpha )\tilde{h}(k-\beta )-N_{-}\tilde{g}%
(k-\alpha )\tilde{h}(k+\beta )) \\
&&+i\omega \tilde{g}(k-\alpha )\tilde{h}(k+\beta )(\tilde{g}(k+\alpha )%
\tilde{h}(k-\beta )+\tilde{g}(k-\alpha )\tilde{h}(k+\beta ))\,.
\end{eqnarray*}
Computation of the $N_{\pm }$ integrals  yields the general forms  
\begin{eqnarray*}
\lambda N_{+}\left( \alpha ,\beta ,k\right)  &=&-i\frac \omega 2\left[ 
\tilde{g}(k-\alpha )\tilde{h}(k+\beta )+\tilde{g}(k+\alpha )\tilde{h}%
(k-\beta )\right] +Q\left( \alpha ,\beta ,k\right) , \\
\lambda N_{-}\left( \alpha ,\beta ,k\right)  &=&i\frac \omega 2\left[ \tilde{%
g}(k-\alpha )\tilde{h}(k+\beta )+\tilde{g}(k+\alpha )\tilde{h}(k-\beta
)\right] +Q\,\left( \alpha ,\beta ,k\right) ,
\end{eqnarray*}
where the function $Q\left( \alpha ,\beta ,k\right) $ is real.

If we now make the additional assumption that our kernel is symmetric, $%
K(x,y)=K(y,x)$, i.e. $g=h$ and $\alpha =\beta $, we obtain $T_{R\rightarrow
L}=T_{L\rightarrow R}$ . It is worthwhile to stress that imposing the
symmetry of the kernel is equivalent to imposing the intertwining condition (%
\ref{T_intertw}), \textit{i.e. }$K_{\mathcal{T}}=K^{\dagger }$, which
ensures the equality of the two transmission coefficients.

A detailed calculation for the case
\[
g(x)=e^{-\gamma|x|}\,,\qquad h(y)=e^{-\delta|y|}\,,
\]
with $\gamma$ and $\delta$ positive numbers, \textit{i.e.} the Yamaguchi
potential, has been presented in Ref.~\cite{CV06}.

Unitarity properties are discussed therein: in particular, when the 
Yamaguchi potential is $\mathcal{PT}$-symmetric with non-zero $\alpha$ and
$\beta$, the characteristics of unitarity breaking are distinctly different
from those of $\mathcal{PT}$-symmetric local potentials discussed in 
Ref.~\cite{Ah04}.

\section{Conclusions}

\label{concl}

In this final section, we try to focus attention on what we believe are the
most original results of our investigation.

Exact $\mathcal{PT}$ invariance has been introduced in the literature as a
condition on the bound-state eigenfunctions of a $\mathcal{PT}$ -symmetric
Hamiltonian. The conclusion is the following: the eigenstates of $H$ should
be eigenstates of $\mathcal{PT}$ , too. In turn, this condition ensures that
the corresponding eigenvalues are real.

As a particular example, we mention the imaginary $\mathcal{PT}$-symmetric
square well studied in Ref.\cite{Zn01}. There, a well defined threshold was
found for the discrete spectrum, separating the regime of exact $\mathcal{PT}
$ symmetry from that of spontaneously broken symmetry. In the case of
scattering, an extension of these results might be ambiguous, in so far as
the continuum can always be labelled with a real energy , $E=k^2$.
Correspondingly, Ref.\cite{Mo05} shows that one can always find $\mathcal{PT}
$-symmetric continuum eigenfunctions of this Hamiltonian, both below and
above the critical potential strength found in Ref.\cite{Zn01}.

In our considerations, we have tried to introduce a specific ''exact'' $%
\mathcal{PT}$ invariance associated with the scattering states of type (\ref
{Psi_1}) and (\ref{Psi_2}). We have called this this condition exact
asymptotic $\mathcal{PT}$ symmetry, which is only and specifically relevant
to scattering; we have shown that this condition forces the $\mathcal{PT}$%
-symmetric potential to be reflectionless ( the above mentioned $\mathcal{PT}
$-symmetric square well does not belong to this class).

While the interest in reflectionless potentials was recently revived in the
frame of supersymmetric quantum mechanics and Darboux transformations\cite
{Ma05}, we stress the fact that the potentials considered in Ref.\cite{Ma05}
are real. Our link of exact asymptotic $\mathcal{PT}$ invariance with
reflectionless complex $\mathcal{PT}$-symmetric potentials should provide
the necessary stimulus to broaden the investigation\cite{An99, ABB05} and
classification of reflectionless potentials so as to include their complex
form, of which we have provided a few examples in the present work.

The other most important topic we have elaborated concerns the delicate
distinction and interplay between hermiticity and time reversal invariance
for non-local potentials and a proper extension of $\mathcal{PT}$ invariance
to this case.

An explicit construction of a solvable separable complex potential has been
presented and worked out in detail in Ref.~\cite{CV06}. 
A particularly notable difference between
local and non-local $\mathcal{PT}$-symmetric potentials is the non-equality
of the two transmission coefficients in the non-local case; they have also a
quite different behaviour in unitarity breaking~\cite{Ah04, CV06}.

An extension of the present work can be envisaged for multi-channel problems%
\cite{BSL03, Zn06, SSFB06}, with the specific aim to formulate a
self-contained and consistent framework to extend the discussion of symmetry
properties to elastic scattering of non-zero spin particles and to inelastic
scattering.

\end{document}